\documentclass[12pt,twocolumn]{IEEEtran}
\usepackage{graphicx}          
\usepackage{amsmath}
\usepackage{tabularx}
\usepackage{amsfonts}
\usepackage{cite}
\usepackage{url}
\usepackage{verbatim}
\usepackage{booktabs}
\usepackage{pifont}
\usepackage{times,epsfig}
\usepackage{makecell}
\usepackage{graphicx}
\usepackage{diagbox}
\usepackage[normalem]{ulem}
\usepackage{multirow}
\usepackage{graphicx}
\usepackage[normalem]{ulem}
\useunder{\uline}{\ul}{}
\usepackage{psfrag}
\usepackage{subfigure}
\usepackage{stfloats}
\usepackage{setspace}
\usepackage{amssymb}
\usepackage{multirow}
\usepackage{colortbl}
\usepackage[justification=centering]{caption}
\usepackage{fancyhdr}
\usepackage{comment}
\usepackage[table,xcdraw]{xcolor}

\usepackage{tikz}
\usetikzlibrary{shapes.geometric, arrows, positioning, fit, calc}

\newcommand{\cmark}{\ding{51}}%
\newcommand{\xmark}{\ding{55}}%
\usepackage{xcolor}

\begin{document}
	
	\pagenumbering{arabic}
	
	\title{Integration of Mixture of Experts and Multimodal Generative AI in Internet of Vehicles: A Survey
	\author{Minrui Xu, Dusit Niyato, \emph{Fellow, IEEE}, Jiawen Kang, Zehui Xiong, \\Abbas Jamalipour, \emph{Fellow, IEEE}, Yuguang Fang, \emph{Fellow, IEEE}, \\Dong In Kim, \emph{Fellow, IEEE}, and Xuemin~(Sherman)~Shen, \emph{Fellow, IEEE}} \vspace{-0.1cm}
 \thanks{M.~Xu and D.~Niyato are with the School of Computer Science and Engineering, Nanyang Technological University, Singapore 639798, Singapore (e-mail: minrui001@e.ntu.edu.sg; dniyato@ntu.edu.sg). J.~Kang is with the School of Automation, Guangdong University of Technology, and Key Laboratory of Intelligent Information Processing and System Integration of IoT, Ministry of Education, Guangzhou 510006, China, and also with Guangdong-HongKong-Macao Joint Laboratory for Smart Discrete Manufacturing, Guangzhou 510006, China (e-mail: kavinkang@gdut.edu.cn). Z.~Xiong is with the Pillar of Information Systems Technology and Design, Singapore University of Technology and Design, Singapore 487372, Singapore (e-mail: zehui\_xiong@sutd.edu.sg). A.~Jamalipour is with the School of Electrical and Information Engineering, University of Sydney, Sydney, NSW 2006, Australia (e-mail: a.jamalipour@ieee.org). Y.~Fang is with Department of Computer Science, City University of Hong Kong, Kowloon, Hong Kong, China. He is a Global STEM Scholar and Chair Professor of Internet of Things (e-mail: my.fang@cityu.edu.hk). D.~I.~Kim is with the Department of Electrical and Computer Engineering, Sungkyunkwan University, Suwon 16419, South Korea (email: dongin@skku.edu). X.~Shen is with the Department of Electrical and Computer Engineering, University of Waterloo, Waterloo, ON N2L 3G1, Canada (e-mail: sshen@uwaterloo.ca).}
}
	\maketitle
	\pagestyle{headings}
	\begin{abstract}

Generative AI (GAI) can enhance the cognitive, reasoning, and planning capabilities of intelligent modules in the Internet of Vehicles (IoV) by synthesizing augmented datasets, completing sensor data, and making sequential decisions. In addition, the mixture of experts (MoE) can enable the distributed and collaborative execution of AI models without performance degradation between connected vehicles. In this survey, we explore the integration of MoE and GAI to enable Artificial General Intelligence in IoV, which can enable the realization of full autonomy for IoV with minimal human supervision and applicability in a wide range of mobility scenarios, including environment monitoring, traffic management, and autonomous driving. In particular, we present the fundamentals of GAI, MoE, and their interplay applications in IoV. Furthermore, we discuss the potential integration of MoE and GAI in IoV, including distributed perception and monitoring, collaborative decision-making and planning, and generative modeling and simulation. Finally, we present several potential research directions for facilitating the integration.
	\end{abstract}

	
\section{Introduction}

\subsection{Background and Motivation}


Artificial General Intelligence (AGI)~\cite{bubeck2023sparks} is believed to enable the full automation in the Internet of Vehicles (IoV)~\cite{zhang2019mobile, naeem2020generative}, which can enable IoV systems to operate autonomously without human supervision and reducing the need for a large human workforce. Specifically, AGI empowers IoV nodes, such as vehicles and roadside units (RSUs), to autonomously and efficiently perform various tasks through their computational capabilities~\cite{chen2023vehicle}. For instance, vehicles can collaborate and form convoys to optimize traffic flow and resource allocation in an AGI-driven adaptive traffic management system. Such a car-following behavior significantly reduces congestion and greatly improves road safety. Therefore, to achieve full automation in IoV, continuous advancements in AI performance are required in specialized tasks such as image recognition, language understanding, and multimodal generative AI (GAI), which can offer new possibilities for various IoV functionalities like collaborative perception and decision-making~\cite{zhang2024smart, qu2024model}.

Fortunately, multimodal GAI~\cite{yan2024forging} can achieve more powerful and secure network performance and diversified services in IoV. In detail, GAI technologies such as generative adversarial networks (GANs)~\cite{kosaraju2019social}, variational autoencoders (VAEs)~\cite{khattar2019mvae}, diffusion models~\cite{xu2023versatile}, and generative transformers~\cite{zhang2024mm} can process and synthesize various modalities of data, including text, images, and point clouds. Specifically, multimodal GAI can achieve sensor data completion, dataset augmentation, and execute complex tasks based on the data collected in IoV. On the one hand, multimodal GAI can provide effective solutions for basic functions in IoV~\cite{tang2019future}, such as resource allocation and vehicular network security. On the other hand, multimodal GAI can enable advanced tasks in IoV. For example, it can process various types of data generated by vehicle sensors using multimodal GAI to enable intelligent capabilities, including safe driving decision-making and traffic prediction~\cite{wen2023road}. Therefore, multimodal AI can not only improve the efficiency of IoV operations and maintenance but also empower IoV with a higher level of intelligence towards full automation of GAI in IoV.

Moreover, the Mixture of Experts (MoE) is an efficient and effective neural network (NN) architecture that divides the parameters of the whole model into multiple expert models with their unique weights~\cite{fedus2022review}. Each expert model excels at handling different aspects of the input data, such as different aspects of IoV operations. In IoV, MoE can facilitate the development of collective intelligence~\cite{galesic2023beyond}, where computations nodes, e.g., vehicles and RSUs, can serve as carriers for one or more expert models~\cite{chen2023vehicle}, trained and operated by local private datasets and algorithms, in the MoE architecture. These computing nodes in IoV can opportunistically perform the necessary computation to fine-tune their local expert models in a distributed manner and then combine them in each node using weighted aggregation to obtain the final output. Through this MoE architecture, GAI models can efficiently compute for better decision-making in a distributed manner in IoV without any performance loss in primary safe driving services~\cite{fedus2022switch, zhenxing2022switch}. In the highly mobile and dynamic IoV~\cite{sun2021applications}, the computation of MoE is even more unpredictable than that over traditional cloud computing, which poses new challenges for its deployment in IoV.

In summary, the integration of MoE and GAI within IoV will bring significant benefits to IoV. However, it is challenging in terms of system implementation and resource management during deployment and execution. In open environments of IoV characterized by high mobility and dynamic topologies, the distributed and collaborative execution of GAI models among computing nodes, e.g., vehicles and RSUs, requires ubiquitous communication and computing resources and seamless synchronization~\cite{sun2021applications}. Specifically, despite the efficiency of MoE, the computational and energy resources required by executing GAI models on vehicles and RSUs are limited for the scalable implementation of MoE systems. 

\subsection{Related Work}

\begin{table*}[t]
\small\centering
\caption{A summary of related surveys.}
\label{tab:algorithm}
\begin{tabular}{|m{.05\textwidth}<{\centering}|p{.6\textwidth}<{\raggedright}|m{.2\textwidth}<{\centering}|}
\hline
Ref. & Description & Topics \\ \hline
\cite{chen2023vehicle} &  Propose the Vehicle as a Service (VaaS) paradigm, which leverages vehicles equipped with Sensing, Communications, Computing, and Intelligence (SCCSI) capabilities to form a service network that enhances the intelligence and service delivery in smart cities. & IoV  \\ \hline
\cite{zhang2019mobile} & Survey the latest developments in edge information systems (EIS) for intelligent IoV, which is crucial for supporting low-latency, localized data acquisition, aggregation, and processing in GAI applications within IoV ecosystem & IoV \\ \hline
\cite{sun2021applications} & Provide a comprehensive survey of the recent developments in applying game theory to solve various challenges in vehicular networks, particularly focusing on issues related to QoS, and security requirements  & IoV \\ \hline
\cite{katare2023survey} & Contribute a comprehensive survey that reviews and compares the latest approximate Edge AI frameworks and publicly available datasets, focusing on energy efficiency for autonomous driving services & IoV/AI \\ \hline
\cite{mao2023green} & Provide a detailed review of energy-efficient design approaches for edge AI systems & AI \\ \hline
\cite{wan2023efficient} & Survey methods to make big AI systems that work with language more resource-friendly
 & GAI \\ \hline
\cite{mcintosh2023google} & Survey the impact of new AI technologies like MoE and multimodal learning on GAI research & GAI/MoE \\ \hline
\cite{wen2023road} & Showcase GPT-4V's advanced understanding and decision-making in autonomous driving scenarios for IoV & IoV/GAI \\ \hline
\cite{zhenxing2022switch} & Provide an overview of datasets that combine vision and language for improved driving and transportation systems & IoV/GAI \\ \hline
\cite{yang2023llm4drive} & Discuss the potential of large AI models in enhancing the understanding and decision-making capabilities of self-driving cars, aiming to replicate human-like behavior & IoV/GAI \\ \hline
\cite{fedus2022review} & Provide a comprehensive review of sparse expert models, which are highly relevant for creating efficient GAI systems due to their ability to decouple parameter count from computation per example & GAI/IoV \\ \hline
Ours & Focus on the efficient training and deployment of scalable multimodal GAI for AGI in IoV & IoV/GAI/MoE \\ \hline
\end{tabular}
\end{table*}

IoV is a subset of the Internet of Things (IoT)~\cite{chen2023rte} in vehicular networks, aiming at enabling vehicles to interact with other vehicles, mobile devices, and roadside infrastructure through vehicle-to-everything (V2X) communications~\cite{chen2023vehicle}. IoV plays a critical role in the development of smart cities by provisioning ubiquitous resources adaptively for supporting various functions, including sensing, communication, computing, and intelligence (SCCSI). The authors in~\cite{chen2023vehicle} introduce the concept of vehicle-as-a-service (VaaS) to leverage vehicles in provisioning smart city services. Towards the next-generation vehicular networks, they outline a novel system architecture for VaaS. Specifically, they review some potential applications of VaaS in smart cities and identify challenges and future research directions. Furthermore, the authors in~\cite{zhang2019mobile} investigate the deployment of mobile edge intelligence for IoV from two perspectives, i.e., vehicle-as-a-client and vehicle-as-a-server. On the one hand, vehicles can access edge resources like data and processing at edge servers of RSUs or base stations (BSs), where edge servers collect and process data from these client vehicles for executing local applications. On the other hand, vehicles can collect data and provision real-time services to serve each other. Although IoV can connect vehicles and edge infrastructure for realizing intelligent transportation systems (ITS), there is a lack of modeling and analyzing strategic interactions among IoV nodes competing or cooperating for limited local and edge resources. The authors in~\cite{sun2021applications} review existing game theory-based solutions to efficient resource allocation, spectrum sharing, routing strategies, and incentive mechanisms to encourage cooperative behavior among vehicles. Beyond traditional optimization and game theory, the authors in \cite{katare2023survey} investigate edge computing and intelligence techniques, which can execute complex AI models for autonomous driving with reduced precision, which significantly lowers the computational burden and energy consumption without substantially compromising the performance of autonomous driving systems. From an energy efficiency point of view, the authors in~\cite{mao2023green} explore how edge AI brings AI processing closer to users to minimize delays, which poses higher energy demands on edge environments.

In GAI, multimodal large language models (LLMs) can be leveraged in language understanding and generation, which is helpful during the interaction between IoV nodes and humans. However, LLMs require enormous resources for training and deployment, which poses significant challenges in the efficient and scalable deployment of LLMs. The authors in~\cite{wan2023efficient} review existing methods to make LLMs more efficient during training and inference, including model design improvement to consume fewer resources during training and inference, training data selection to reduce convergence time, and leveraging special software to improve inference speed of LLMs. Furthermore, the authors in \cite{mcintosh2023google} explore the impact of MoE, multimodal learning, and AGI on GAI research. Specifically, MoE uses many small-scale NNs to tackle inference tasks collaboratively, which is achieved by using a gating network to select the optimal expert network for each task. Combining text, images, and other data types, multimodal LLMs can be used for understanding and generating complex information. Based on multimodal LLMs, the authors in~\cite{wen2023road} investigate the potential of GAI in autonomous driving which relies on integrating perception, decision-making, and control systems. While traditional methods struggle with complex driving environments and predicting other road users' actions, GPT-4V(ision) is a new Visual Language Model (VLM) tested for autonomous driving, which excels at scene understanding and causal reasoning over current systems. For multimodal LLMs in ITS, the authors in \cite{zhou2023vision} survey current models and datasets for training, fine-tuning, and inference for vehicular applications. Furthermore, the authors in \cite{yang2023llm4drive} review the current research status of leveraging LLMs for enabling autonomous driving from the angle of enhancement in understanding, reasoning, and decision-making in existing driving systems. The authors in \cite{fedus2022review} review existing sparse expert models, including architectures like MoE and Switch transformers for efficient training and executing of large AI models.
\begin{figure*}
    \centering
    \includegraphics[width=1\linewidth]{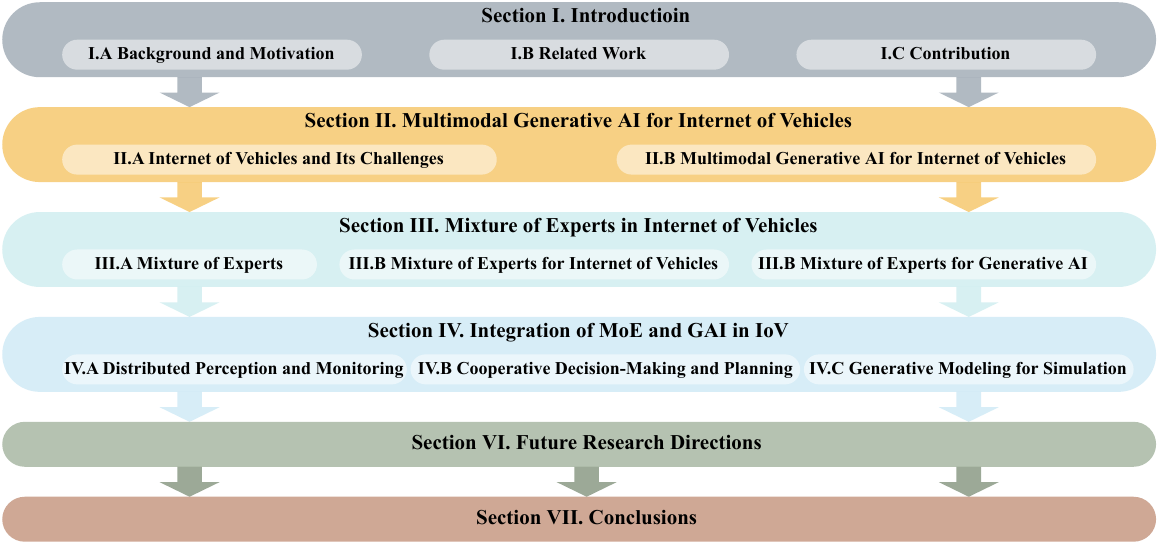}
    \caption{The outline of this survey.}
    \label{fig:outline}
\end{figure*}
\subsection{Contribution}

In this paper, we focus on the efficient training and deployment of scalable multimodal GAI for AGI in IoV. Aiming at emphasizing the importance of integrating MoE and GAI to advance AGI in IoV systems, we elucidate how this integration not only effectively enhances the intelligence of individual vehicles but also significantly raises the collective intelligence of vehicular networks. Through an in-depth exploration of the interplay and synergies between MoE and GAI, we provide a comprehensive review of existing efforts of this combination's potential to revolutionize IoV systems. 

Moreover, we accentuate how IoV systems can act as a facilitator for the advanced integration of MoE and GAI. The distinct characteristics of IoV, characterized by its expansive network, diverse data streams, and imperative for real-time operational capabilities, present a feasible platform for refining and enhancing the collaborative mechanisms between MoE and GAI. This convergence is anticipated to not only augment the capabilities of IoV systems but also to pave the way for innovative applications in intelligent transportation and other vertical fields such as smart cities and healthcare industries.

Our main contributions can be summarized as follows:
\begin{itemize}
    \item We introduce the utilization of multimodal GAI for IoV, which can enhance traffic operations, driving safety, and social traffic intelligence. Specifically, GAI technologies can be leveraged for intelligent transmission scheduling and spectrum management in IoV, addressing the challenges of channel variability and spectrum variability.
    \item To utilize MoE in IoV, we present the MoE architecture and discuss the existing utilization status of MoE in IoV systems for intelligent traffic management and autonomous driving.
    \item We highlight the potential of integration of MoE and GAI as a significant advancement for AGI in IoV, enhancing both individual vehicle intelligence and the collective intelligence of vehicular networks.
    \item We provide several future research directions, by identifying privacy-preserving collaborative inference as a key area for future research, emphasizing the need for balancing computational efficiency with privacy in the data-constraint IoV systems.
\end{itemize}

The outline of this survey is presented in Fig. \ref{fig:outline}.

\section{Multimodal Generative AI for Internet of Vehicles}







\subsection{Internet of Vehicles and Its Challenges}

By utilizing V2X communications~\cite{zhang2019mobile, naeem2020generative, chen2023vehicle}, IoV systems can improve the availability of in-network resources and capabilities, boost the efficiency of traffic operations, enhance driving safety, increase the intelligence of social traffic services~\cite{tang2019future}, and expand the service provisioning for other vertical industries~\cite{chen2023vehicle}. In detail, IoV systems consist of vehicles and roadside communications and computing infrastructure, enabling the collection and processing of traffic data from onboard sensors and computing servers and then transporting information for intelligent decision-making and control~\cite{xu2017internet}. For transportation, this integration enables the development of innovative applications such as autonomous driving and platooning, while also improving safety and energy efficiency. IoV creates a distributed network where sensors and interconnected devices, e.g., base stations, RSUs, and capability-empowered vehicles, exchange crucial information and conduct needed computing, facilitating real-time applications and efficient implementation in IoV systems~\cite{ye2023accuracy}. However, ensuring smooth communication, computing, and interoperability among interconnected nodes in IoV remains challenging.

In IoV, high mobility and dynamic topology of IoV nodes~\cite{sun2021applications} can cause fast and unpredictable changes in the network and result in frequent disconnections and reconnections as vehicles move, leading to unstable communication links and increased packet loss. Moreover, high-speed vehicles on highways may experience rapid changes in network topologies, necessitating swift handover decision-making~\cite{biswas2023autonomous}. To address these dynamic conditions, robust handover schemes and efficient routing protocols are required to adapt to the high-speed movement of vehicles~\cite{zardari2022adaptive}. Furthermore, ensuring Quality of Service (QoS) becomes complex under high mobility conditions~\cite{jin2022mobility}, as the network must dynamically allocate resources to maintain service levels despite variations in network density and topology.

\begin{figure*}
    \centering
    \includegraphics[width=0.8\linewidth]{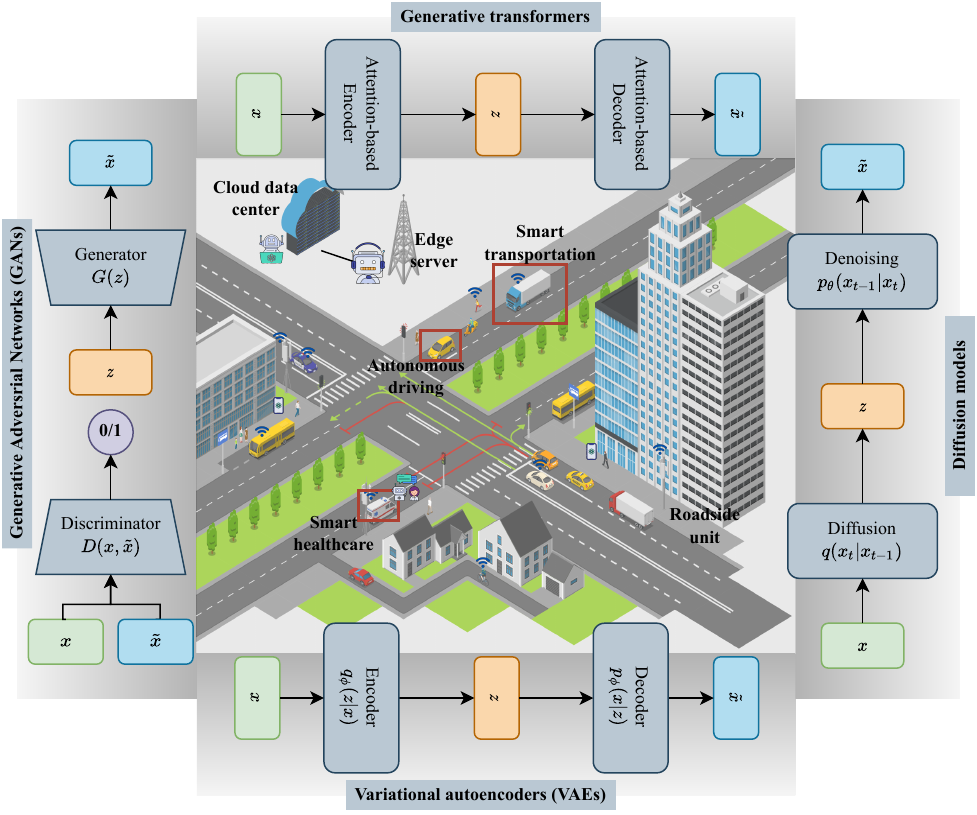}
    \caption{Internet of Vehicles supported by different generative AI technologies.}
    \label{fig:IoVGAI}
\end{figure*}

In addition, with the growing number of vehicles and mobile devices, scarce resources such as congested spectrum or access media, limited bandwidth, computational capacity, storage, and energy have become a major concern~\cite{chen2023vehicle}. Balancing the allocation of these resources becomes crucial, as conflicts among different requirements arise, creating a competition for resource allocation. For instance, resource-intensive applications like high-definition (HD) video streaming or real-time data analytics for traffic management intensify the competition for bandwidth and computational resources~\cite{ince2022real}, while adopting strong security algorithms will introduce significant overhead in terms of communication delay, high computing power, energy consumption, or space utilization~\cite{liu2021blockchain}. Efficient and effective distributed execution in IoV relies on carefully managing these limited resources. Technologies such as edge computing, where computational tasks are offloaded to nearby edge nodes, can mitigate these resource constraints.

Furthermore, cooperation among vehicles and other infrastructure under dynamic topology becomes more challenging in IoV, especially under dynamic environmental changes. The performance and reliability of communications in vehicular networks can be affected by various factors such as channel fading and weather conditions like rain~\cite{khairnar2013performance}. For example, cooperative vehicular communication in urban scenarios, where buildings may cause signal reflections and fading, is essential for maintaining connectivity. In dense traffic scenarios, frequent information broadcasting can lead to congestion, demanding for cooperative measures. Additionally, the open nature of communication channels in IoV systems requires cooperation to address security threats and ensure network integrity against adversary attacks. In the literature, existing cooperative solutions among vehicles and roadside infrastructure, like cooperative beamforming, could enhance signal strength and security~\cite{gu2020control}.

\subsection{Multimodal Generative AI for Internet of Vehicles}

\begin{table*}[!t]
\small\centering
\caption{Summary of GAI for resource allocation and network security in IoV.}
\label{tab:GAIIOV}
\begin{tabular}{|c|c|m{.1\textwidth}<{\centering}|c|c|}
\hline
Service & Paper & Task & GAI & Insight \\ \hline
\multirow{15}{.08\textwidth}{\centering Resource Allocation} & \cite{naeem2020generative} &Transmission scheduling & GAN & \begin{tabular}[c]{p{.48\textwidth}} \vspace{-2mm} {\color{green}\cmark} GAN is used for approximating the action-value distribution in transmission scheduling. \\ \vspace{-2mm} {\color{green}\cmark} GAN-Scheduling can intelligently schedule packet transmission under extreme network conditions. \\ \vspace{-2mm} {\color{red}\xmark} Unable to generalize across diverse network conditions due to limited datasets \end{tabular} \\ \cline{2-5} 
 & \cite{lin2021tulvcan} & Spectrum management & GAN & \begin{tabular}[c]{p{.48\textwidth}} \vspace{-2mm}{\color{green}\cmark} The goal is to design a sensing matrix and reconstruction algorithm for spectrum recovery.\\ \vspace{-2mm} {\color{green}\cmark} High-quality reconstructed spectrum enables the identification of unused frequency bands and complex spectrum-sharing designs. \\ \vspace{-2mm} {\color{red}\xmark} The robustness of the model is not enough against a wide range of spectrum utilization scenarios, including those with sparse or non-uniform spectral activity\end{tabular} \\ \cline{2-5} 
 & \cite{falahatraftar2021conditional} & Network slicing & GAN & \begin{tabular}[c]{p{.48\textwidth}} \vspace{-2mm}{\color{green}\cmark} CGANs can generate synthetic data that closely resemble real network scenarios\\ \vspace{-2mm} {\color{red}\xmark} The reliance on a centralized SDN architecture could introduce a single point of failure, potentially impacting the reliability of the network\end{tabular} \\ \hline
\multirow{25}{.08\textwidth}{\centering Vehicular Network Security} & \cite{seo2018gids, chen2023gan} & Intrusion detection & GAN & \begin{tabular}[c]{p{.48\textwidth}} \vspace{-2mm} {\color{green}\cmark} GIDS can detect unknown attacks using only normal data.\\ \vspace{-2mm} {\color{red}\xmark} Requires frequent updates when in-vehicle environments change\end{tabular} \\ \cline{2-5} 
 & \cite{qiu2022unsupervised} & Driving anomaly detection & GAN & \begin{tabular}[c]{p{.48\textwidth}} \vspace{-2mm} {\color{green}\cmark} GAN models to extract feature embeddings from the penultimate layer of the discriminator for each modality.\\ \vspace{-2mm} {\color{red}\xmark} The GAN-based approach is not scalable and cannot incorporate more modalities if needed.\end{tabular} \\ \cline{2-5} 
 & \cite{monshizadeh2021improving} & Traffic anomaly detection & VAE & \begin{tabular}[c]{p{.48\textwidth}} \vspace{-2mm} {\color{green}\cmark} CVAE introduces labels of traffic packets into a latent space for better learning.\\ \vspace{-2mm} {\color{red}\xmark} The solution may improve the detection rate for a specific dataset but varies for another dataset and each type of attack.\end{tabular} \\ \cline{2-5} 
 & \cite{aslam2023vae} & Traffic video anomaly detection & VAE & \begin{tabular}[c]{p{.48\textwidth}} \vspace{-2mm} {\color{green}\cmark} The A-VAE is trained in an unsupervised manner on regular video sequences.\\ \vspace{-2mm} {\color{red}\xmark} The unsupervised nature of the training also means that the model may not perform optimally if the training data is not representative of the diversity of normal traffic patterns.\end{tabular} \\ \cline{2-5} 
 & \cite{li2022mfvt} & Anomaly traffic detection & Transformer & \begin{tabular}[c]{p{.48\textwidth}} \vspace{-2mm} {\color{green}\cmark} The transformer encoder's multi-head attention mechanism allows the model to focus on different parts of the input data simultaneously.\\ \vspace{-2mm} {\color{red}\xmark} Transformers may struggle with very long-range dependencies because the self-attention mechanism, despite its global reach, might not always effectively capture the importance of distant token relationships.\end{tabular} \\ \cline{2-5} 
 & \cite{nwafor2022canbert} & Intrusion detection & Transformer & \begin{tabular}[c]{p{.48\textwidth}} \vspace{-2mm} {\color{green}\cmark} The use of efficient transformer models optimized for smaller datasets and memory-constrained environments can mitigate these limitations.\\ \vspace{-2mm} {\color{red}\xmark} The high computational requirement and processing time are involved in pre-training.\end{tabular} \\ \hline
\end{tabular}%

\end{table*}

As illustrated in Fig. \ref{fig:IoVGAI}, multimodal GAI can enhance various applications in IoV including resource allocation and vehicular network security. For instance, the authors in \cite{jagadish2022conditional} use conditional VAE networks to predict the future paths of traffic agents in autonomous vehicular applications, which is crucial for understanding and anticipating the behavior of other vehicles to ensure safe maneuvering. In the literature, GAI approaches that consider past trajectories and traffic scenes encoded in a latent space have been proposed for multimodal predictions with fewer parameters and better performance than baseline models, even when predicting across different time horizons. In general, a few key issues must be considered.

\subsubsection{Resource Allocation}

Resource allocation in IoV systems is a critical task that involves effectively distributing and managing resources, including bandwidth, computing units, and energy, to facilitate communication and processing/computing for IoV nodes~\cite{chen2023tasks}. Utilizing dynamic resource allocation techniques based on GAI, such as GANs, VAE, and Transformers, IoV systems can allocate resources adaptively according to dynamic demands and network conditions. These advanced techniques consider factors, such as traffic conditions, network availability, and QoS requirements, to optimize resource allocation. A primary objective of resource allocation in IoV is to enhance communication reliability, minimize latency, and improve overall system performance~\cite{raza2021task}. Specifically, GANs are employed to address underestimation issues in noisy and dynamic IoV systems, thereby enhancing transmission scheduling. In this context, the transmission scheduling of vehicles emerges as a pivotal factor for ensuring timely and reliable communication among vehicles and infrastructure. For instance, the authors in \cite{naeem2020generative} introduce a novel GAN-scheduling framework to augment the flexibility and intelligence of vehicular networks. This framework is particularly motivated by the need for efficient resource management and scheduling in the dynamic and complex environments of IoV systems~\cite{li2021adaptive}. Employing GANs as an agent within the Software-defined Networking (SDN) control plane, the system state is assessed through various network parameters, with the agent tasked with storing transitions and updating network weights. The proposed approach leverages cognitive mechanisms to adapt to changing network conditions and optimize scheduling decisions.

Moreover, the authors in \cite{lin2021tulvcan} propose the Compression and Reconstruction Network (CRNet), a deep learning-based framework tailored for terahertz (THz) spectrum sensing and reconstruction in vehicular networks. This innovative CRNet method integrates a structured sensing matrix with a two-phase reconstruction process, employing a one-layer CNN for compression and advanced reconstruction modules for accurate spectrum recovery. Specifically, the authors demonstrate that CRNet excels beyond GAN-based algorithms, particularly in its robustness across different compression rates, a vital aspect for efficient vehicular communication. Similarly, the authors in \cite{falahatraftar2021conditional} delve into the evolving nature of Heterogeneous Vehicular Networks (HetVNETs) and the subsequent challenges for network slicing within the context of 5G technology. Responding to the need for efficient resource allocation and congestion management for HetVNETs, they propose a novel hybrid architecture that merges SDN and Network Functions Virtualization (NFV) with Conditional GANs. The experimental results demonstrate that different batch sizes affect performance and using a batch size of 40 is recommended when adopting the proposed CGAN.

\subsubsection{Vehicular Security}

Unfortunately, high mobility complicates the implementation of security measures, as the constantly changing network topology makes it difficult to maintain a secure and consistent network-wide policy~\cite{huang2022blockchain,hou2023data}. Moreover, implementing encryption protocols in such a dynamic environment is also challenging due to the need for frequent key exchanges and authentication~\cite{huang2020dapa}. GANs are used to enhance vehicular network security by detecting anomalies and unauthorized access. In the context of vehicular network security, GANs utilize a two-network architecture consisting of a generator network and a discriminator network. The generator network produces synthetic data that mimics normal network behavior, while the discriminator network distinguishes between real and synthetic network data. By training the GAN solely on normal data, the model can identify anomalies and intrusions, including unknown attacks not present in the training dataset. This approach is particularly effective in dynamic environments such as vehicular networks, where traditional signature-based detection systems may struggle to detect new threats. In IoV, the research is motivated by the security vulnerabilities of the Controller Area Network (CAN) bus. Despite its efficiency, the CAN bus lacks robust security features to protect against intrusions. To address this gap, the authors in \cite{seo2018gids} propose an Intrusion Detection System (IDS) that utilizes GANs to detect both known and unknown attacks on in-vehicle networks. The IDS model employs a dual-discriminator approach. The first discriminator is trained on known attacks using actual vehicular data, while the second discriminator is trained adversarially with the generator to identify unknown attacks by distinguishing real and fake CAN images generated from random noise. By achieving at least 98\% accuracy in four types of attacks, this method shows promise in improving the accuracy of intrusion detection systems for in-vehicle networks, which is crucial when considering the potential safety implications for drivers. The authors propose in \cite{chen2023gan} design an anomaly detection algorithm based on GANs to address the imbalance between normal and attack traffic in vehicular networks. This algorithm is trained using only normal traffic data and employs the Gini index for data dimensionality reduction and normalization functions to improve training speed. The authors in \cite{qiu2022unsupervised} discuss the important requirement for improved safety in vehicular applications by using GAI to identify driving anomalies. The motivation behind this is the potential to decrease accidents and fatalities by detecting irregular driving patterns early on. The proposed approach involves training conditional GANs separately for each modality, such as CAN-Bus data, physiological signals, and environmental information. The system's ability to distinguish anomalies is significantly enhanced by incorporating environmental modalities, especially with pedestrian distances.

In vehicular network security, VAEs have emerged as a key tool for enhancing vehicular network security by providing effective anomaly detection and intrusion detection capabilities. VAEs excel at learning the normal patterns and behaviors of vehicular network traffic, thereby enabling them to detect any deviations or anomalies that may indicate security threats. In a pioneering effort, the authors in \cite{monshizadeh2021improving} offer an innovative design that blends a Conditional VAE (CVAE) with a Random Forest (RF) classifier. This combination addresses the challenges of data generalization and overfitting in network traffic anomaly detection, a crucial aspect in the context of vehicular networks where ensuring cybersecurity and safety is paramount. The integration of CVAE's generative capabilities with RF's classification prowess leads to improved performance in attack classification when compared to conventional feature selection methods. Building on the theme of network security, the authors in \cite{kim2023anomaly} introduce a method for detecting anomalies in in-vehicle networks through the innovative use of multiple LSTM-Autoencoder models. This approach is in response to the need to enhance the security of the CAN protocol, which currently lacks robust security measures and is susceptible to different types of attacks. GAI is applied to deliver a solution that is both accurate and effective in vehicular applications, considering various features such as transmission intervals and payload value changes,. Further extending the scope of vehicular network security, the authors in \cite{aslam2023vae} leverage an Attention-based VAE (A-VAE) for detecting anomalies in traffic videos. Driven by the imperative for intelligent traffic monitoring systems to ensure safety and security, this method combines a GAI approach with a novel blend of 2D-CNN, Bi-LSTM layers, and an attention mechanism. The resultant system demonstrates high accuracy and real-time performance over challenging traffic datasets.

Adding to the diversity of approaches, the authors in \cite{nguyen2023transformer} present a transformer-based attention network as a revolutionary intrusion detection system for CAN buses in vehicles. Motivated by the pressing need to improve vehicular security against malicious attacks due to the CAN protocol's inherent vulnerabilities, this method employs self-attention mechanisms for efficient multi-class classification of attacks, including replay attacks, and utilizes transfer learning to boost performance on small datasets from different car models. In a similar vein, the authors in \cite{li2022mfvt} develop the MFVT model, a groundbreaking anomaly traffic detection method that integrates a feature fusion network with a vision transformer architecture. Aimed at combating the increasing cyber-attacks on vehicular networks and ensuring their safe operation, this method, although not explicitly mentioning GAI, focuses on enhancing intrusion detection systems by effectively identifying suspicious behaviors in network traffic, thus playing a crucial role in safeguarding the network's integrity and reducing economic losses. 

Additionally, the authors in \cite{li2020detecting} introduce a novel anomaly detection method for intelligent vehicle charging and station power supply systems. This approach is underscored by the growing necessity for secure and reliable smart charging infrastructure in transportation electrification. The proposed Multi-Head Attention (MHA) model, leveraging the Google Transformer encoder architecture, excels at identifying anomalies by capturing the inherent correlations in traffic generated by Industrial Control Systems (ICS) with high accuracy, thus outperforming traditional and CNN-based models in real-world power ICS testbed. Complementing these approaches, the authors in \cite{nwafor2022canbert} present CANBERT, a language-based intrusion detection model for in-vehicle networks. Addressing the critical requirement for robust security against network attacks such as replay, fuzzing, and denial of service in vehicular applications, this method harnesses the deep semantic understanding capabilities of transformer-based BERT models to detect malicious attacks on the CAN bus with high precision and accuracy.


\textbf{Lessons Learned:} To address various challenges in IoV, including high mobility, cooperation under dynamic environments, and limited resources, GAI-based methods can be introduced to improve resource utilization efficiency and ensure vehicular network security. Specifically, GAI-based methods can augment the datasets for training AI-based methods. In this way, these enhanced AI-based methods can be trained on large-scale datasets and thus achieve higher performance with GAI. However, there are still several limitations to these GAI-based methods. For instance, the GAI methods usually require additional resources for training data generation and discrimination models, which cannot be accomplished in real-time to adopt dynamic mobile environments. In addition, GAI-based methods cannot guarantee robustness in various scenarios, which might consist of non-uniform activity. To address these issues, attention-based methods, i.e., transformers, are leveraged to improve the performance of these models during the optimization in smaller datasets and memory-constrained environments. The efficient execution of AI-based generative methods is crucial to spark AGI in IoV.

\section{Mixture of Experts in Internet of Vehicles}



\subsection{Mixture of Experts}
\begin{figure*}
    \centering
    \includegraphics[width=1\linewidth]{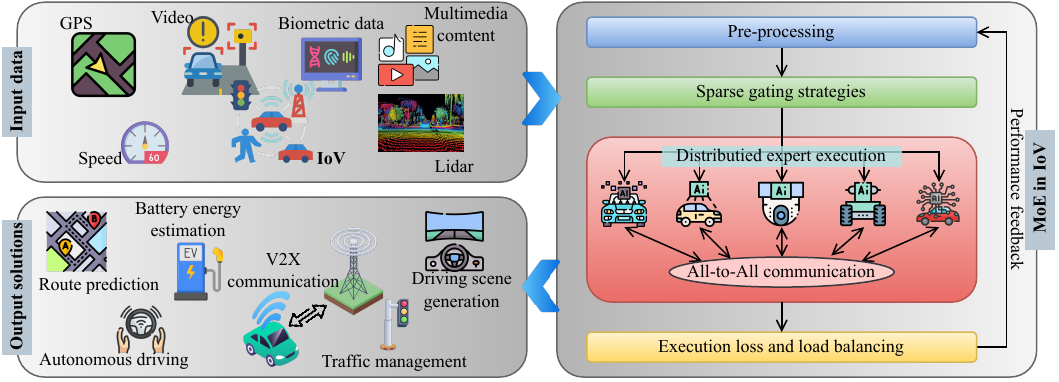}
    \caption{Mixture-of-experts (MoE) Architecture and its applications in IoV.}
    \label{fig:IoVMoE}
\end{figure*}

The concept of the MoE architecture can be traced back to ensemble learning, where multiple models (base learners) are trained to solve the same problem and their prediction results are combined simply, such as through voting or averaging~\cite{fedus2022review}. The main objective of ensemble learning is to improve prediction performance by reducing overfitting and enhancing generalization capability. While multimodal GAI can provide basic functionalities in IoV such as efficient resource allocation and protection of in-vehicle network security, its ability to simultaneously process data from sources such as text, images, and speech, the scale of multimodal AI models needs to increase in terms of parameters, complexity, and scale to incorporate more knowledge and possess stronger reasoning capabilities. To achieve scalable GAI for AGI in IoV, the MoE architecture can be utilized to scale the parameters of GAI models without a proportional increase in computational cost~\cite{fedus2022review}. By deploying expert models in computing nodes within IoV, the MoE architecture can leverage the strengths of different expert models in different vehicles or RSUs, allowing each expert to focus on different regions of the input space, e.g., road segments, thereby improving the performance of the final results. As illustrated in \ref{fig:IoVMoE}, the implementation details of MoE include four main building blocks, i.e., sparse gating strategies, distributed expert execution, all-to-all communication, and load balancing.


\subsubsection{Sparse Gating Strategies}

The MoE architecture consists of a set of $N$ ``expert networks" $\mathcal{E} = \{E_1, \ldots, E_n, \ldots, E_N\}$, and a ``gating network" $G$ whose output is a sparse $n$-dimensional vector~\cite{shazeer2017outrageously}. Each expert model is an NN that runs on computing nodes in IoV, each with its own set of parameters. Without loss of generality, all the expert models that can collaborate on an IoV computing node can process inputs of the same dimension and produce outputs of the same dimension.

The MoE is based on the diverse data sets which all contribute collectively to the generalization of overall learning performance. Given input $x$ for the MoE layer, the output $y$ of the MoE layer can be written as follows:
\begin{equation}
    y = \sum_{i=1}^{N} p_i(x) E_i(x),
\end{equation}
where gate-value $p_i$ for expert $i$ is obtained from the gating network $G$, which is determined by the gating strategies. The computation cost is determined by the sparsity of the gate-value $p_i$, where no computation is performed for executing $E_i(x)$ is needed when $p_i(x)=0$.

The gating strategies are key components of all the MoE architecture, determining which experts to allocate. For instance, the gating network of the top-$k$ gating function~\cite{fedus2022switch} has a trainable variable $W_r$ which computes the logits $h(x)=W_r\cdot x$, which are normalized via a softmax distribution over the $N$ experts, i.e.,
\begin{equation}
    p_i(x) = \frac{e^{h(x)_i}}{\sum_{j=1}^{N}e^{h(x)_j}}.
\end{equation}

The computation cost for the gating functions alone is at least $O(TN)$ for all $T$ tokens in the input batch given $N$ experts. However, in our study, $T$ is in the order of millions and $N$ is in the order of thousands, a sequential implementation of the gating functions would keep most of the computational resources idle most of the time. Therefore, we need an efficient parallel implementation of the gating functions to leverage many devices.

\subsubsection{Distributed Expert Execution}

After the gating network determines the allocation decision for sparse expert models, the computing nodes that have been allocated these expert models utilize computation resources to execute them locally. These expert models are distributed among IoV nodes. In the distributed MoE implementation, the expert capacity $c$ is leveraged to characterize the computation cost of each expert~\cite{fedus2022switch}, which is set by evenly dividing the number of tokens in the batch across the number of experts as
\begin{equation}
    c = \sigma (T/N),
\end{equation}
where $\sigma$ denotes the capacity factor. If the capacity ratio is greater than 1.0, additional buffers can be provided to handle the uneven distribution of tokens among experts. If too many tokens are allocated to an expert (discarded tokens will be mentioned later), the computation will be skipped and the token representation will be directly passed to the next layer through the remaining connections. However, increasing expert capacity also has its drawbacks as higher capacity values can lead to unnecessary computation and memory usage. Although larger capacity factor values can improve quality, they can also increase communication, memory, and computational costs. Efficient implementation of the all-to-all primitive, along with adjusting gating algorithms, such as reducing the capacity factor, can help mitigate the additional communication costs associated with sparse expert algorithms.

\subsubsection{All-to-all Communications}
Assigning inputs to experts is typically implemented as an all-to-all communication primitive, where each computing node sends data to all other nodes. In many gating algorithms (but not all), the forward pass incurs two all-to-all communication costs, while the backward pass incurs an additional two all-to-all communication costs. For example, gating algorithms including the BASE layer require four all-to-all communications in the forward pass and four all-to-all communications in the backward pass~\cite{lewis2021base}. Due to the additional all-to-all communication costs involved in implementing gating, networks with slower interconnect speeds may find that fewer expert layers are optimal in terms of time to achieve a certain quality level.

Considering the computational, storage, and communication resources of IoV, GAI can greatly assist network managers in quickly determining the optimal settings without the need for costly trial and error. This communication pattern is necessary for globally determining the token allocation of each expert and is also a significant factor in the communication overhead of sparse expert models. Efficient implementation of all-to-all communication and modification gate algorithms, such as reducing the capacity factor, helps reduce these costs. The capacity factor itself directly affects the communication cost by adjusting the batch size of the experts. Effectively managing these communication costs is crucial, as the communication cost may become a bottleneck as the number of experts in the distributed system increases~\cite{lepikhin2020gshard}.

\subsubsection{Load Balancing}

It would be the best if the MoE layer could sparsely activate the expert with the given token. The simplest method is to select the best expert based on the soft maximum probability distribution. However, it is well known that this approach leads to the problem of an imbalanced training load, where most of the tokens seen during training are assigned to a few busy experts, resulting in a large accumulation of input buffers for those few experts, which slows down the training speed for other experts, while many other experts do not receive sufficient training at all. To address these issues, Switch Transformers introduce an auxiliary load-balancing loss to be added to the total modal loss during training~\cite{fedus2022switch}, which is computed as
\begin{equation}
    loss = \alpha  N  \sum_{i=1}^{N} f_i P_i,
\end{equation}
where $\alpha$ is a multiplicative
coefficient, $f_i = \frac{1}{T}\sum_{x\in\mathcal{B}} \mathbf{1}\{\arg\max p(x) =i\}$ is the fraction of tokens dispatched to expert $i$ and $P_i = \frac{1}{T}\sum_{x\in \mathcal{B}}p_i(x)$ is the fraction of the routing probability allocated for expert $i$ across all tokens in the batch $\mathcal{B}$. The auxiliary loss can encourage uniform gating since it is minimized under a uniform distribution.

\subsubsection{Communication-efficient MoE Architecture}

Due to the limited communication and computing resources in IoV, the communication-efficient MoE architecture is promising to be adopted to leverage limited bandwidth to perform all-to-all communication during the synchronization among expert models. Furthermore, to efficiently utilize all the computing resources in IoV, MoE architecture that can balance the load among IoV nodes is also required. Here we introduce several communication-efficient MoE architectures to address these issues.

\textbf{Switch transformers} \cite{fedus2022switch} are a class of models that scale the parameter count of a Transformer architecture while maintaining a constant computational cost per example. They utilize a sparsely activated model design, which means that different parameters are selected for each incoming example, allowing the model to have a significantly larger number of parameters without a proportional increase in computational demands. The architecture replaces the dense feed-forward network in the Transformer with a sparse Switch feed-forward network layer, which operates independently on the tokens in the sequence. By designing models based on T5-Base and T5-Large, the Switch Transformer achieves up to 7x increases in pre-training speed with the same computational resources. This indicates that the architecture can handle a larger number of parameters without a linear increase in computational cost. Furthermore, the Switch Transformer models show gains in multilingual settings, outperforming the mT5-Base version across all 101 languages tested.

\textbf{GLaM} \cite{du2022glam} or Generalist Language Model, is a family of language models that utilize a sparsely activated MoE architecture to efficiently scale model capacity. The largest GLaM model boasts 1.2 trillion parameters, which is approximately 7 times larger than GPT-3, yet it requires only one-third of the energy for training and half the computation flops for inference compared to GPT-3. Despite its efficiency, GLaM demonstrates superior performance across a range of zero, one, and few-shot NLP tasks. Specifically, the largest GLaM model, with 1.2 trillion parameters, demonstrates this superior performance while consuming only one-third of the energy used to train GPT-3 and requiring half the computation flops for inference. This efficiency is attributed to the sparsely activated MoE architecture, which activates only a small fraction of the total parameters for each prediction, allowing the model to scale effectively by increasing the size or the number of experts in the MoE layer.

\subsection{Mixture of Experts for Internet of Vehicles}

\begin{table*}[]
\small\centering
\caption{Summary of MoE-based approaches in IoV.}
\label{tab:MoEIoV}
\begin{tabular}{|c|c|m{.12\textwidth}<{\centering}|m{.1\textwidth}<{\centering}|m{.5\textwidth}<{\raggedright}|}
\hline
Service & Paper & Task & AI model & MoE-based Approach \\ \hline
\multirow{15}{.1\textwidth}{\centering Intelligent traffic management} & \cite{shen2023traffic} & Traffic prediction & Transformer & MoE transformer  uses an adaptive attention mechanism to focus on relevant spatiotemporal relationships \\ \cline{2-5} 
 & \cite{wang2022st} & Traffic prediction & ResNet, LSTM, CNN & ST-ExpertNet is an explainable framework for traffic prediction that leverages a MoE-based approach to specialize different flow patterns \\ \cline{2-5} 
 & \cite{petersen2022data} & Battery energy estimation & MLP & MORE uses expert models trained on link-wise data aggregated by GPS information from a navigation provider, which distinguishes five different road types \\ \cline{2-5} 
 & \cite{yuan2023temporal} & Vehicle trajectory prediction & CNN & TM-MoE consists of a shared layer for extracting temporal features using a Temporal Convolutional Network (TCN), an expert layer with a gating mechanism for sequence memory and filtering, and a fully connected layer for integrating and outputting predictions \\ \cline{2-5} 
 & \cite{fraser2023deep} & Drone trajectory prediction & LSTM & DMoE uses expert regression models to predict extreme bounds of trajectories for specific intent classes, like perimeter flight, point-to-point flights, package delivery, and area mapping \\ \hline
\multirow{12}{.1\textwidth}{\centering Autonomous driving} & \cite{fang2020multi} & Multimodal perception & CNN & The gating network of the proposed MoE-based approach, trained through a multi-stage procedure, determines the most informative input for the driving task, addressing computational complexity and overfitting. \\ \cline{2-5} 
 & \cite{pini2023safe} & Safe autonomous driving & Transformer & SafePathNet aims to improve on-road safety by predicting and planning with a MoE and has been validated through extensive simulation and real-world deployment on urban public roads. \\ \cline{2-5} 
 & \cite{morra2023mixo} & Visual Odometry & CNN & The MoE approach combines outputs from different experts, representing single-camera odometry algorithms and inertial odometry, to estimate the vehicle's trajectory. \\ \cline{2-5} 
 & \cite{john2018estimation} & Steering angle estimation & LSTM & The proposed MoE-based approach uses an LSTM network to model the regression function with multiple experts. \\ \cline{2-5} 
 & \cite{enzweiler2011multilevel} & Pedestrian classification & MLP & The multilevel MoE framework for pedestrian classification, motivated by the need for reliable pedestrian detection in vehicular applications to enhance safety in urban traffic \\ \hline
\end{tabular}
\end{table*}

In the literature, several MoE-based approaches are proposed to enable intelligent traffic management and autonomous driving in IoV.

\subsubsection{Intelligent Traffic Management}

\begin{figure}
    \centering
    \includegraphics[width=1\linewidth]{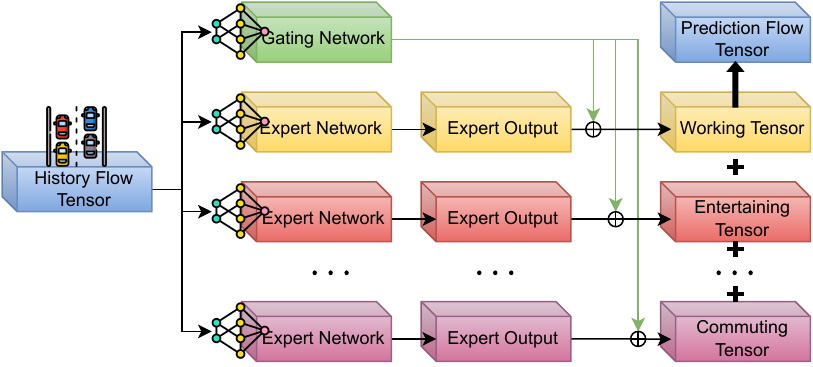}
    \caption{The workflow of traffic management leveraging the MoE architecture for future traffic prediction.}
    \label{fig:traffic}
\end{figure}

Traffic flow prediction aims to forecast future traffic flow on cities or highways~\cite{wang2022sfl}. This problem is complex due to dynamic and nonlinear traffic patterns influenced by various factors like daily trends, weather conditions, and unexpected events. Traditional time series models are not effective in handling the intricate spatiotemporal dependencies in traffic data, resulting in poor prediction performance\cite{wang2022traffic}. To address this issue, the mixture-of-expert Transformer model proposed in~\cite{shen2023traffic} integrates the concept of expert specialization and adaptive attention mechanisms, which can capture complex dependencies in traffic data by utilizing multiple expert models and a gating network to weigh their contributions based on the input data. It also uses an adaptive attention mechanism to focus on relevant spatiotemporal relationships, leading to superior performance and computational efficiency compared to traditional time-series models and existing transformer-based methods. Moreover, the traffic prediction problem involves forecasting vehicle or crowd flow within a city, which is important for urban traffic management and public safety, as it helps anticipate demands on transportation infrastructure~\cite{yang2022urban}. Specifically, traffic flow is influenced by various factors, such as recent time intervals, daily periodicity, and weekly trends, while traditional approaches struggle with the mixed state of flow patterns, which are influenced by the functional distribution of city areas. Fortunately, the MoE architecture allows for the decomposition of the traffic prediction task into subtasks, with each expert model specializing in a distinct flow pattern. As illustrated in Fig.~\ref{fig:traffic}, the authors in~\cite{wang2022st} introduce ST-ExpertNet, an explainable framework for traffic prediction that leverages an MoE-based approach to specialize different flow patterns, motivated by the need to address the mixed state of citywide flow patterns such as commuting and commercial flows. The proposed approach involves training multiple experts with a gating network to ensure each expert is responsible for predicting specific flow patterns, aiming to enhance interpretability and performance in traffic prediction.

The problem of energy estimation for battery electric vehicles (BEVs) is related to accurately predicting the remaining range of the vehicle based on its battery capacity, which is a crucial factor in combating range anxiety among users\cite{maity2023data}. Range anxiety is the fear that the vehicle battery may run out of energy before reaching the destination or a charging station, which is a significant barrier to the further uptake of BEVs. The complexity of this problem stems from the multitude of factors that influence the energy consumption of a BEV, including driving style, road topology, weather, and traffic conditions. Nevertheless, traditional approaches cannot adequately account for these variables, resulting in inaccurate range estimates. To address this issue, the authors in \cite{petersen2022data} develop a data-driven approach using machine learning and ensemble learning techniques, specifically the MoE method to train specialized NNs on different road types and combine their predictions to improve the accuracy of energy consumption estimations for BEVs. The methodology, named Mixture of Road Energy Experts (MORE), uses expert models trained on link-wise data aggregated by GPS information from a navigation provider, which distinguishes five different road types. The expert models used are Multilayer Perceptron (MLP) architectures, chosen for their high accuracy in time series estimation and their ability to serve as a baseline for investigating the MORE approach. The input to the MLPs includes various statistical features, with a correlation-based feature selection process used to deduce six relevant features that significantly impact energy consumption. This approach has demonstrated a reduction in the root mean squared error (RMSE) of energy estimation by approximately 7.5\% compared to a monolithic model, indicating a significant improvement in the predictive performance.

Vehicle trajectory prediction for autonomous vehicles and driver assistance systems is essential~\cite{lu2022vehicle}, allowing for the anticipation of future vehicle positions for safe and efficient driving decisions. The Temporal Multi-task MoE (TM-MoE) model proposed in~\cite{yuan2023temporal} is a sophisticated approach that integrates the prediction of vehicle trajectory and driving intention, acknowledging the interrelated nature of these tasks. The model consists of a shared layer for extracting temporal features using a Temporal Convolutional Network (TCN), an expert layer with a gating mechanism for sequence memory and filtering, and a fully connected layer for integrating and outputting predictions. Compared to other models like LSTM, CNN, CNN-LSTM, TCN, and GRU, the TM-MoE model has shown better performance, especially for input sequence lengths of 3s, 6s, and 9s. A sensitivity analysis also confirms that incorporating driving intentions significantly improves the accuracy of vehicle trajectory prediction. The model's effectiveness is validated using the CitySim dataset and includes innovative elements such as a lane line reconstruction method to reduce measurement errors in the dataset. For drone trajectory prediction~\cite{yang2022study}, the authors in \cite{fraser2023deep} propose the deep MoE (DMoE), which significantly advances the classification and prediction of drone trajectory intent using non-cooperative radar data. The DMoE framework integrates a high-level intent classifier, called BCLSTM-A, which is excellent at recognizing patterns in trajectory data. It predicts the intent class of a drone's trajectory, helping understand its purpose and future actions. In addition, the DMoE uses expert regression models to predict extreme bounds of trajectories for specific intent classes, like perimeter flight, point-to-point flights, package delivery, and area mapping. These expert models are combined with the probabilities from the intent classifier, improving the prediction of trajectory bounds. This approach addresses noise and uncertainty in radar track data, providing a more reliable estimation of a drone's future location for airspace management and conflict avoidance. The architecture's performance is validated against various deep learning models, showing its superior accuracy in classification and prediction tasks with different flight profiles and time windows.

\subsubsection{Autonomous Driving}

Autonomous driving refers to the technology that allows vehicles to operate without human intervention by perceiving the environment, making decisions, and controlling the vehicle's movement~\cite{li2023towards}. It relies on a combination of sensors, such as cameras and LiDAR, to gather information about its surroundings, and advanced algorithms to process this data and make driving decisions~\cite{danapal2020sensor}. For multimodal sensing data processing, the authors in~\cite{fang2020multi} design a multi-modal expert network tailored for autonomous driving, motivated by the need to enhance perception and reliability through sensor fusion while addressing computational complexity and overfitting challenges. The MoE-based approach is used in this network architecture for autonomous driving, which selects the most relevant sensor input at each step, acting as a MoE model. There are three camera inputs and one LiDAR input. The gating mechanism, trained through a multi-stage procedure, determines the most informative input for the driving task, addressing computational complexity and overfitting. The network's output ensures a balanced utilization of all experts, preventing attention from partial. The architecture is demonstrated on a truck with multiple sensors, handling failures and maintaining predictive power for steering commands. This approach is critical for autonomous vehicles in complex driving environments.

\begin{figure}
    \centering
    \includegraphics[width=1\linewidth]{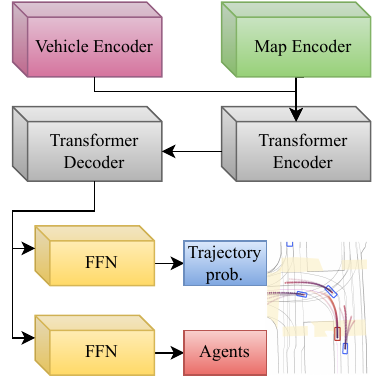}
    \caption{SafePathNet can predict future trajectories and road agents using MoE architecture based on the input scene.}
    \label{fig:safeAV}
\end{figure}

Safe real-world autonomous driving refers to self-driving vehicles (SDVs) navigating public roads without human intervention~\cite{acerbo2021safe, vitelli2022safetynet, wylde2012safe}. It involves end-to-end AI systems that learn complex behaviors instead of relying on handcrafted rules, which includes an NN architecture that models future trajectories for SDV and other road agents, selecting the safest planning trajectory based on predicted probabilities and minimizing a safety cost. As illustrated in Fig.~\ref{fig:safeAV}, the authors in~\cite{pini2023safe} propose SafePathNet aiming to improve on-road safety by predicting and planning with a MoE and has been validated through extensive simulations and real-world deployment on urban public roads. The goal is to ensure SDV operates safely and comfortably under various driving scenarios. The MoE approach models the distribution of future trajectories for SDV and other IoV nodes. It uses multiple NNs to predict different trajectory options and their probabilities collaboratively. These predictions are then used to improve driving safety by penalizing plans that could cause collisions. The MoE approach is an important part of the model for prediction and planning, which is tested extensively in a realistic simulator and deployed on SDVs under real-world conditions.

Visual Odometry (VO) estimates an agent's trajectory by analyzing sequential camera images to determine relative motion between frames~\cite{cho2023dynamic}. VO algorithms can be categorized based on camera configurations, such as monocular, stereo, or RGB-D, and the keypoint extraction method, such as direct, feature-based, or hybrid. Feature-based VO matches feature descriptors across frames and balances accuracy and computational efficiency optimally. The authors in \cite{morra2023mixo} utilize the MoE framework in multi-camera VO for autonomous systems, specifically with the MIXO technique. The MoE approach combines outputs from different experts, representing single-camera odometry algorithms and inertial odometry, to estimate the vehicle's trajectory. The gating function in the MoE framework selects and weighs each expert's contribution based on the current condition. This optimally fuses incremental yaw rates estimated by each camera, using the scale estimated by the same inertial odometry model. The gating function predicts the optimal yaw rate combination, while the MIXO technique predicts the optimal weighting scheme based on the previous yaw rate.

The estimation of steering angle and collision avoidance in autonomous driving is addressed by proposing a monocular camera-based method that utilizes a second-order particle filtering algorithm to robustly estimate steering angles for varying driving patterns~\cite{paponpen2022implementation}. For robust steering angle estimation in autonomous driving, the authors in~\cite{john2018estimation} develop a MoE framework utilized within an end-to-end learning framework for the proposal distribution in a particle filtering algorithm. The proposed MoE-based approach uses an LSTM network to model the regression function with multiple experts. It leverages a gating network to determine the weights of the outputs from different experts. This approach models the second-order dynamics of the proposal distribution, which is important for accurately tracking the steering angle in different driving patterns. The LSTM within the MoE framework addresses the vanishing gradient problem and improves accuracy in estimation. The performance of this MoE framework is validated by comparing it with a straightforward LSTM-based regressor, showing its robustness and accuracy in estimating steering angles under different road scenes and driving behaviors.

In IoV, pedestrian classification is the process of identifying and categorizing pedestrians in various environments using computer vision techniques. The authors in \cite{enzweiler2011multilevel} present a novel multilevel MoE framework for pedestrian classification, motivated by the need for reliable pedestrian detection in vehicular applications to enhance safety in urban traffic. The proposed framework combines multiple features and cues, such as image intensity, dense depth, and dense flow. It uses expert classifiers like MLPs and linSVMs to reduce false positives while maintaining detection rates. The MoE framework improves pedestrian classification by combining features and cues at different levels: pose, modality, and feature. Shape cues based on Chamfer shape matching provide priors for pedestrian views at the pose level. At the modality level, individual classifiers are used for image intensity, dense depth, and dense flow. The feature-level MoE combines histograms of oriented gradients (HOGs) and local binary patterns (LBPs) to enhance performance across all modalities, with the most significant improvement in the intensity modality. The multilevel MoE framework integrates these classifiers and outperforms other fusion approaches and thus can effectively address the high-dimensional space challenge without overfitting.

\subsection{Mixture of Experts for Generative AI}

In addition, the introduction of MoE to GAI is expected to be a viable solution to the efficient training and deployment of GAI models, which are usually composed of billions of parameters.

\textbf{GAN:} The introduction of MoE to GAN can address the limited representation capacity and uninformative learning signals obtained from the discriminator in GANs. For instance, the authors in~\cite{gurumurthy2017deligan} propose DeLiGAN, which is a novel GAN architecture designed to generate diverse images from limited datasets by reparameterizing the latent space as a mixture model, which is learned alongside the GAN parameters to enhance sample diversity. The MoE approach is introduced to GAN through the DeLiGAN framework, where the latent space is reparameterized as a mixture model, specifically a mixture-of-Gaussians model. This reparameterization allows the model to capture a more diverse set of features and modes from limited data by learning a set of Gaussian distributions, each representing different ``experts" within the latent space. The parameters of these Gaussian components, including their means and variances, are learned during the training process alongside the traditional GAN parameters. This model enhances the generative capability of the network, enabling it to produce a wider variety of samples that better represent the underlying data distribution, even when the available training data is scarce. Its ability to approximate complex distributions is the key to this improvement, as it allows the network to model the high-probability regions in the latent space more effectively. The DeLiGAN framework demonstrates the practical benefits of this approach across various data modalities, including handwritten digits, objects, and hand-drawn sketches. In addition, the authors in~\cite{park2018megan} introduce a novel GAN model, namely, MEGAN, which incorporates a mixture of expert models to encourage generators within the GAN to learn different modalities present in the data. This is achieved by employing multiple generator networks that specialize in generating images with specific subsets of modalities, and a gating network that selects the appropriate generator for a given condition. In addition, the MEGAN utilizes the Gumbel-Softmax reparameterization trick along with a developed regularization for load-balancing, which further stabilizes the training of MEGAN. Furthermore, the authors in~\cite{chai2023improved} demonstrate that an MoE approach can significantly enhance the representation capacity of the generator in language GANs. This is achieved by incorporating multiple cooperative experts within the generator network, which attempt to jointly generate high-quality sentences through their interactions, thereby providing a more expressive and diverse generation capability to compete with the discriminator. The MoE generator allows for a richer representation of the data distribution, which is particularly beneficial for the complex patterns inherent in human language. Additionally, the MoE paradigm can contribute to the diversity of the generated samples, as multiple experts enrich the representation capacity of generators. This approach is orthogonal to other improvements on generators in language GANs and represents a significant shift from the conventional single-expert generator architectures.

\textbf{VAE:} For conditional computation and task decomposition of VAEs, the authors in~\cite{yu2021mixture} develop a novel anomaly detection (AD) framework that leverages a MoE ensemble with two convolutional VAEs and a convolution network, named MEx-CVAEC. In MEx-CVAEC, the network explicitly learns the underlying manifold of a group of similar objects for anomaly detection (AD) by employing an encoder-decoder-encoder (EDE) pipeline with VAE as the core element in each expert structure. The gating network, which is a convolutional autoencoder (CAE), learns to use a subset of experts conditioned on the input examples, thereby optimizing the expert structures for efficient data characterization based on the manifold of the latent space. The proposed MEx-CVAEC model captures the differences of latent spaces by encoding data and modeling it as a mixture of low-dimensional nonlinear manifolds, with the gating network assembling experts with different weights to make full use of the experts. This approach is the first to suggest a mixture of CVAEs-based models for AD, and it utilizes the advantages of combining reconstruction error and multi-manifold latent space detection. Moreover, the MoE-Sim-VAE proposed in~\cite{kopf2021mixture} leverages a Gaussian mixture model in the latent space to represent the similarities between data points, which is encouraged through the optimization process. This approach enables the model to perform both clustering and generation tasks with high efficacy and efficiency, as demonstrated on various datasets including MNIST, single-cell RNA-sequencing, and mass cytometry measurements.

For multi-modal generative models, the authors in \cite{shi2019variational} propose a variational MoE autoencoder (MMVAE) framework, which is demonstrated to satisfy four key criteria for successful learning: implicit latent decomposition, coherent joint generation, coherent cross-generation, and improved model learning through multi-modal integration. This MoE approach allows the model to learn from multiple modalities by effectively combining the expertise of individual unimodal encoders to form a joint representation. The MoE framework addresses the challenge of cross-modal generation by enabling the model to generate data in one modality given data in another, without the need for all modalities to be present at all times. This is particularly useful for handling missing data at test time and for improving the coherence of joint and cross-modal generations. The proposed MMVAE leverages this MoE approach to satisfy the criteria for successful multi-modal generative model learning. For image compression tasks, the authors in \cite{fleig2023edge} present a novel approach to estimate Steered-Mixtures-of-Experts (SMoE) parameters through the design of an innovative Autoencoder network that incorporates embedded SMoE decoder capabilities. This strategy marks a significant advancement in run-time performance, achieving encoder run-time savings by a factor of 500 to 1000 compared to previous works, without compromising the quality of the reconstructed images. The Autoencoder is designed to map pixel blocks directly to model parameters for compression, which aligns with recent trends in the algorithm ``unfolding," while remaining fully compatible with the established SMoE framework. The proposed method not only accelerates the encoding process to real-time capabilities but also enhances rate-distortion performance. To extend the SMoE proposed in \cite{fleig2023edge}, the authors in \cite{fleig2023steered} leverage SMoE in real-time image modeling and denoising to improve the reconstruction quality of complex and noisy images while drastically reducing the computational time required compared to gradient descent optimized SMoE models.  

\textbf{Diffusion Model:} Diffusion models based on MoE have several advantages for machine learning, which can effectively act as filters for different frequency ranges at each time-step noise, allowing for specialized operations for high and low frequencies. This is achieved by employing denoisers with specialized architectures tailored to the operations required at each time-step interval. The authors in \cite{Luo2023ImageSV} introduce ERNIE-ViLG 2.0, a large-scale Chinese text-to-image diffusion model that addresses the challenges of image fidelity and text relevancy in text-to-image generation. The MoE concept is introduced to diffusion models through the Mixture-of-Denoising-Experts (MoDE) mechanism, which involves dividing the denoising process into different stages, each handled by a specialized expert network. The primary motivation behind MoDE is the recognition that tasks at different timesteps during the denoising process are distinct and that using the same set of parameters for all tasks might lead to suboptimal performance. Therefore, the timesteps are uniformly divided into blocks, with each block assigned to a specific denoising expert. This allows each expert to focus on the characteristics of the denoising steps within its block, improving the model's performance by tailoring the denoising process to the noise ratio of the input at each stage. The introduction of MoDE enables ERNIE-ViLG 2.0 to scale up the parameters of the diffusion model without increasing inference time, as only one expert network is activated at each step. In addition, the authors in \cite{Luo2023ImageSV} achieve image super-resolution (SR) using diffusion models. Firstly, it identifies the issue of information loss within the latent diffusion model used for image SR and proposes a frequency-compensated decoder complemented by a refinement network to infuse more high-frequency details into the reconstructed images, thereby enhancing overall image quality. Secondly, the paper introduces the concept of Sampling-Space MoE (SS-MoE) to enlarge the diffusion model for image SR, which allows for enhanced high-resolution image processing without necessitating a substantial increase in training and inference resources, optimizing the efficiency significantly.

\textbf{NERF:} The authors in \cite{Mi2023SwitchNeRFLS} present Switch-NeRF, a novel framework for large-scale Neural Radiance Fields (NeRF) that introduces a learning-based scene decomposition strategy. The key contributions include the development of a gating network that efficiently dispatches 3D points to different NeRF sub-networks, which can be jointly optimized with the sub-networks using a Sparsely Gated MoE design. To achieve scene decomposition with MoE, the Switch-NeRF framework introduces a learnable gating network that dynamically dispatches 3D points to different NeRF sub-networks, each acting as an expert for a particular scene partition. This gating network is optimized in conjunction with the NeRF sub-networks using a Sparsely Gated MoE design, allowing for end-to-end learning of scene decomposition. The MoE-based Switch-NeRF model ensures that the outputs from different sub-networks are fused in a learnable manner within the unified framework, effectively maintaining the consistency of the entire scene. The design of the gating network is such that it can be trained without human intervention, and it does not require any priors of the 3D scene shape or the distribution of scene images, making it a generic solution for large-scale scenes. This design is carefully implemented and optimized to achieve high-fidelity scene reconstruction and efficient computation, establishing state-of-the-art performance on several large-scale datasets.

\section{Integration of MoE and GAI in IoV}

\subsection{Distributed Perception and Monitoring}

IoV enables autonomous vehicles to perceive their surroundings by exchanging sensor data with nearby vehicles, improving perception accuracy beyond their sensing range~\cite{zhang2019mobile}. During the distributed perception and monitoring, GAI can be leveraged in super-resolution (SR) of the images captured by the vehicle's sensors, which is essential for accurate object detection and navigation, ensuring road safety. Specifically, SR techniques are employed to improve the quality of images that are often degraded due to factors like distance, weather conditions, and sensor limitations. GANs have emerged as a powerful method for SR, providing high perceptual quality in enhanced images. For instance, the SD-GAN model focuses on generating sharper images with excellent perceptual quality, despite lower Peak Signal-to-Noise Ratio (PSNR) values and some distortion, which can be mitigated by combining pixel-wise Charbonnier loss with the adversarial loss function~\cite{xia2022pluralistic}. In addition, the CE-GAN approach is proposed in \cite{jiang2022gan}, which is a novel method that leverages LiDAR point cloud data to enhance and restore images from in-vehicle cameras, addressing the challenge of image quality degradation due to adverse weather or obstructions. The CE-GAN framework is designed to overcome issues such as over-smoothing in damaged areas, which can cause a loss of detailed texture when using GAN networks for image repair. By using a combination of loss functions, including LossL1, LossSA, and LossBlurriness, the CE-GAN aims to produce high-quality images that retain detail and texture, outperforming conventional methods like Cycle-GAN. This enhancement is critical for IoV, as it directly impacts the vehicle's ability to perceive and interact with its environment, thereby influencing the safety of autonomous driving. To train and validate perception systems for autonomous vehicles in IoV, the authors in~\cite{xu2021reliability} leverage a combination of real-world data and synthetic data generated by GANs. 

To address the challenge of training and validating vision-based perception systems for autonomous vehicles, motivated by the need to handle rare and unseen scenarios in vehicular applications. GANs are used to generate and augment realistic data without domain shift, specifically employing models like CycleGAN~\cite{zhu2017unpaired} for day-to-night image translation and pix2pixHD for video manipulation to create long-tail situations such as sharp cut-ins by other vehicles. The possible extensions of these methods in an MoE could involve integrating GAN-generated data into a broader framework where multiple specialized models or ``experts" are trained on both real and synthetic data, enhancing the robustness and reliability of autonomous systems by covering a wider range of driving scenarios and conditions. Learning latent dynamics for autonomous driving tasks within IoV involves creating models to predict a vehicle's actions based on sensor data. To address the challenge of predicting vehicle behavior in complex environments, the authors in~\cite{pak2022carnet} propose a Combined dynAmic autoencodeR Network (CARNet) that integrates autoencoders and recurrent NNs to learn current and future latent representations from high-dimensional sensor data, offering a more parameter-efficient model that outperforms existing architectures like World Models (WM) in both imitation and reinforcement learning tasks. By incorporating MoE in this method, different specialized models (experts) are combined to handle various aspects of the driving environment, potentially improving the robustness and generalization of the autonomous system across diverse scenarios. In this regard, the authors in~\cite{xia2022pluralistic} focus on pluralistic image completion using a novel end-to-end probabilistic method that leverages a Gaussian Mixture Model (GMM) to ensure diversity in the generated images, which is crucial for applications where multiple plausible completions are desirable. The proposed method, which is based on a probabilistic graph model, divides the image completion process into sub-procedures and optimizes the inherent parameters of GMM during training to adaptively meet the diversity requirements of the task.

Sensor data enhancement in IoV can improve the accuracy and quality of data collected from sensors in vehicles, which is important for ensuring the safety of drivers and pedestrians, as well as for optimizing the performance of IoV applications. GPS sensors are integral components of autonomous vehicles within IoV, providing critical positioning data that enables these cars to navigate safely to their desired destinations. Motivated by the need for accurate and reliable positioning data crucial for safe navigation, the authors in \cite{karlsson2018data} propose a data-driven generative model to employ autoregressive processes and Gaussian Mixture Models to stochastically simulate GPS sensor errors, capturing the complex error characteristics and variations over time. To ensure the safety and reliability of Advanced Driver Assistance Systems (ADAS) and autonomous driving (AD), the authors in \cite{arnelid2019recurrent} propose an improved Recurrent Conditional Generative Adversarial Network (RC-GAN) that employs RNNs with LSTM units in both the generator and discriminator, conditioned on environmental sequences to generate synthetic sensor errors with long-term temporal correlations. This approach allows for capturing spatial and temporal dependencies, enabling more realistic simulations that are crucial for virtual testing in the development of safe autonomous vehicles.

Modeling human driving behavior is crucial for the development of autonomous driving systems and for enhancing traffic simulation accuracy. This modeling is approached as a sequential decision-making problem under uncertainty, characterized by continuous state and action spaces, non-linearity, stochasticity, and an unknown cost function. Motivated by the need for accurate simulation and safety validation of autonomous driving systems, the authors in \cite{bhattacharyya2022modeling} propose Generative Adversarial Imitation Learning (GAIL) and its extensions, namely, PS-GAIL, for multi-agent interaction, RAIL for domain-specific knowledge integration, and Burn-InfoGAIL for latent variability disentanglement—to learn neural driving policies from real-world data. To extend \cite{bhattacharyya2022modeling}, the authors in \cite{kuefler2017imitating} train recurrent NN policies, demonstrating that these models can replicate emergent human driving behaviors and maintain control over long time horizons in realistic highway simulations.

In IoV, vehicles are equipped with various sensors and communication devices that enable them to collect and share information with infrastructure components. This information can include vehicle speed, direction, location, and data about the surrounding environment, such as road conditions, traffic signals, and obstacles. The authors in~\cite{li2023variational} propose an inter-instance variational auto-encoder (IIns-VAE), for concurrent distance estimation and environmental identification from wireless signals, which encapsulates both distance and environmental features in latent variables. This methodology bridges the gap between statistical inference and deep learning, leveraging the efficiency of deep learning for approximating complex distributions and the interpretability of statistical techniques to avoid overfitting. Realistic semantic segmentation is crucial for autonomous urban driving within IoV, as it enables vehicles to interpret and understand their surroundings at a pixel level, distinguishing between different objects such as cars, pedestrians, and road signs. Motivated by the need for accurate real-time perception in IoV, the authors in \cite{yi2021improving} propose a generative adversarial network (GAN) based method, specifically PGE-GAN, which utilizes an ensembling framework to enhance the generalization of synthetic-to-realistic translation, thereby improving the performance of semantic segmentation in vehicular applications. The proposed method leverages co-training classifiers and various loss functions to train the ensemble of GANs, aiming to reduce the domain gap between synthetic and real-world data. Therefore, possible extensions of this work could involve integrating the proposed GAN approach with an MoE system, which could further refine the model's predictions by combining the strengths of multiple expert models, potentially leading to even more robust and accurate semantic segmentation for autonomous driving applications.

Driving scene understanding within IoV is a critical aspect of autonomous vehicle (AV) technology, where semantic segmentation is a key vehicular technology that assigns pixel-level labels to images by enabling AVs to accurately locate and understand traffic objects in their surroundings. To address the challenge of domain adaptation in vehicular applications, the authors in \cite{hua2023domain} propose a novel uncertainty-aware ensemble-based GAN framework, named UE2D-GAN, which enhances the generalization of semantic segmentation models by training on diversified optimization objectives, training iterations, and network initializations. The proposed method outperforms other advanced methods in terms of mean Intersection-over-Union (mIoU) on benchmark datasets. Possible extensions of this work could involve integrating the UE2D-GAN framework into a mixture of expert systems, where the uncertainty-aware fusion strategy could be used to weigh the contributions of different expert models, potentially improving the robustness and reliability of driving scene understanding in complex IoV scenarios. Furthermore, the integration of Visual-Language Models (VLMs) like GPT-4V into autonomous driving represents a significant advancement. The authors in~\cite{zhou2023vision} investigate the potential of GPT-4V, a GAI model, in autonomous driving, motivated by the need for improved scene understanding and decision-making in vehicular applications. The paper presents a methodical evaluation of GPT-4V's capabilities, from basic scenario comprehension to advanced causal reasoning, highlighting its proficiency in interpreting complex driving environments and predicting the behavior of other road users.

\subsection{Cooperative Decision-Making and Planning}

Cooperative decision-making and planning in IoV is a crucial aspect of the successful integration of highly automated and autonomous vehicles on public roads~\cite{varga2023cooperative}. The challenge lies in addressing interactions with vulnerable road users, especially at low driving speeds. Several papers propose algorithms for synchronizing local plans in cooperative distributed decision-making, enabling parallel decision-making and achieving feasible and globally consistent decisions~\cite{kloock2023coordinated}. These algorithms consist of two iterative steps: local planning and global synchronization. In the local planning step, agents compute local decisions, while global synchronization ensures consistency across agents. The globally synchronized plans act as reference decisions for the next iteration. The convergence to globally feasible decisions depends on the feasibility of the coupling topology. The proposed algorithms have been evaluated using car-like robots and have shown the achievement of globally feasible decisions. 

\subsubsection{Trajectory Prediction and Route Planning}

Trajectory prediction in IoV for autonomous driving is a critical function that ensures the safety and efficiency of self-driving vehicles by forecasting the future motion of surrounding traffic agents, such as cars, buses, bicycles, and pedestrians. The authors in \cite{zhang2022ai} address the challenge of ensuring safety in autonomous driving by proposing the Attention-based Interaction-aware Trajectory Prediction (AI-TP) model, which leverages GAI to predict the trajectories of traffic agents with high accuracy and low inference time. The AI-TP model employs Graph Attention Networks (GAT) to capture the complex interactions among traffic agents and Convolutional Gated Recurrent Units (ConvGRU) for generating future trajectory predictions, demonstrating superior performance over existing methods. Moreover, the authors in \cite{fu2021progrpgan} propose the Progressive Route Planning Generative Adversarial Network (ProgRPGAN), which utilizes a novel GAN framework to generate realistic paths with increasing map resolution. Specifically, ProgRPGAN generates realistic paths by employing a progressive training approach where a sequence of generator and discriminator pairs are trained on grid maps of increasing resolutions, eventually refining the path on the actual road network. The generators in ProgRPGAN produce sequences of grid cells or road segments that represent the path, while the discriminators classify these paths as real or generated. The framework utilizes Long Short-Term Memory (LSTM) networks for both generators and discriminators, which allows for the sequential prediction of the next cell or intersection on the path. To enhance the training efficiency and stability, ProgRPGAN transfers parameters from previous-level generators and discriminators to subsequent ones. Additionally, it pre-trains embeddings of grid cells and intersections to capture the network topology and external factors, such as road segment types, which facilitates effective model learning. This method not only speeds up the training process but also ensures the generation of diverse and realistic routes for the same query. 

The authors in \cite{ocampo2023improving} propose a Multimodal Generative Model for Path Planning that could leverage the principles of Wasserstein GANs (WGANs) with Gradient Penalty (GP) to approximate the distribution of free configuration spaces, which is essential for navigating through unknown environments with obstacles. This approach would involve conditioning the WGAN-GP with a Variational Auto-Encoder (VAE) in a continuous latent space to handle multimodal datasets, such as those encountered in IoV scenarios where sensor data from various modalities must be integrated for effective path planning. The VAE's role would be to encode the conditioning from the obstacles' image scenario of the WGAN-GP generator, enabling the generation of configuration states that closely match those encountered in previously seen scenarios and explore new configurations. This model could be extended to higher-dimensional configuration spaces, making it a versatile and effective method for sampling-based path planning and configuration space (CS) reconstruction in the dynamic and complex environments typical of IoV applications. The use of local critics in the WGAN models could further improve performance by ensuring that the model converges to a better distribution and enhances its ability to generate valid paths. This advanced generative modeling approach could significantly enhance the performance of IoV systems by reducing the computational cost of planning while still producing optimal paths. 

The authors in \cite{he2023diffusion} propose the Multi-Task Diffusion Model (MTDIFF), a novel diffusion-based method that leverages Transformer backbones and prompt learning to effectively plan and synthesize data in multi-task reinforcement learning (RL) settings, motivated by the need for generalist agents capable of handling diverse tasks in complex environments such as those encountered in vehicular applications. Moreover, the authors in \cite{huang2023diffusion} present SceneDiffuser, a novel diffusion-based generative model for 3D scene understanding that integrates generation, optimization, and planning into a unified framework, demonstrating significant improvements over previous models in tasks such as human pose and motion generation, dexterous grasp generation, and path planning for 3D navigation. SceneDiffuser can be incorporated into a mixture of expert systems by serving as a specialized expert for generating and optimizing trajectories in 3D environments. In this way, SceneDiffuser's ability to generate physically plausible and diverse trajectories conditioned on complex 3D scenes would complement other expert models that might be specialized in different aspects of the task at hand, such as object recognition or semantic scene understanding.

\subsubsection{Car-following and Platooning}

Car-following models are crucial for the longitudinal control of connected and autonomous vehicles (CAVs) within IoV. These models are designed to simulate and predict the behavior of a following vehicle based on the state of a lead vehicle, such as its speed, acceleration, and relative distance. In the context of IoV, car-following models are particularly important as they enable CAVs to adapt to dynamic and mixed traffic conditions where human-driven vehicles (HVs) and CAVs coexist. To address this issue, the authors in \cite{ma2023physics} propose a novel hybrid model, the Physics-Informed Conditional Generative Adversarial Network (PICGAN), which combines physics-based and data-driven approaches without the need for explicit weighting parameters, leveraging GAI to improve multi-step car-following modeling in mixed traffic flow scenarios. Motivated by the necessity to improve traffic flow and safety in mixed-traffic environments in \cite{ma2023physics}, the authors in \cite{ma2023application} propose a novel multi-step car-following model using a Conditional GAN that captures the decision-making process of human drivers and generates actions that closely resemble human driving behavior, which is crucial for the integration of CAVs in mixed traffic flows.

\subsubsection{Human-in-the-loop Decision Making}

In IoV, LLMs such as GPT-4 are integrated to facilitate human-like interactions within autonomous vehicles, transforming the communication paradigm from rigid commands to natural, intuitive conversations. The authors in \cite{cui2024drive} propose a novel framework aimed at enhancing the decision-making processes of autonomous vehicles by integrating LLMs, motivated by the need for human-centric design and advanced AI capabilities in vehicular applications. The proposed method involves leveraging the natural language processing and contextual understanding of LLMs, such as ChatGPT-4, to improve interaction, reasoning, and adaptability in autonomous driving scenarios, with a focus on personalized assistance and continuous learning. The authors in \cite{sha2023languagempc} propose a novel approach to autonomous driving by integrating LLMs to enhance decision-making in vehicular applications, motivated by the need for systems that can intuitively align with user preferences for driving behavior and handle complex scenarios. The proposed method involves a chain-of-thought framework that employs LLMs for logical reasoning and high-level decision-making, which is then translated into mathematical representations to guide a Model Predictive Control (MPC) bottom-level controller, demonstrating significant improvements over existing learning-based and optimization-based methods.

\begin{figure}
    \centering
    \includegraphics{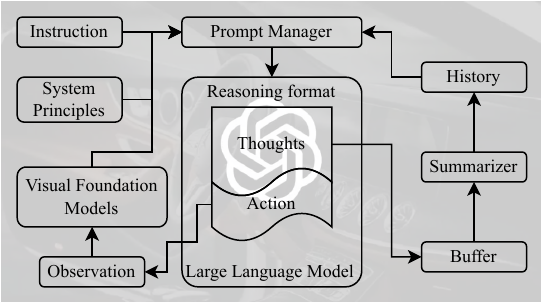}
    \caption{The architecture of NavGPT~\cite{zhou2023navgpt}.}
    \label{fig:navgpt}
\end{figure}

Considering the issues of dataset bias, overfitting, and uninterpretability in pertaining foundation models for autonomous driving, the authors in~\cite{wen2023dilu} propose the DiLu framework, which integrates interactive environments, driver agents, and a memory component to instill human-like knowledge-driven capabilities into autonomous driving systems. The DiLu framework employs LLMs for reasoning and reflection, leveraging their common-sense knowledge and emergent abilities to make decisions and evolve continuously, demonstrating superior generalization over reinforcement learning-based methods. As illustrated in Fig. \ref{fig:navgpt}, the authors in~\cite{zhou2023navgpt} introduce NavGPT, an LLM-based navigation agent, motivated by the potential of LLMs trained on vast language data to exhibit significant reasoning abilities, which is crucial for developing universal embodied agents for applications such as autonomous vehicular navigation. The proposed method leverages the reasoning capabilities of GPT models to perform zero-shot sequential action prediction in vision-and-language navigation tasks, integrating high-level planning, commonsense knowledge, and adaptability to dynamic environments, which could be pivotal in complex vehicular applications where understanding and navigating real-world scenes are essential.

\subsection{Generative Modeling for Simulation}

\subsubsection{Traffic Simulation}
In IoV, generating diverse and realistic traffic scenarios is essential for evaluating the AI safety of autonomous driving systems, which is important for the comprehensive evaluation of vehicular applications. Therefore, the authors in \cite{feng2023trafficgen} propose TrafficGen, which leverages a data-driven GAI approach that synthesizes both the initial states of traffic vehicles and their long-term trajectories from real-world datasets, using an encoder-decoder neural architecture to capture the complexity of traffic dynamics. Specifically, TrafficGen generates realistic traffic scenarios by employing a novel encoder-decoder architecture that first encodes a scene snapshot using a vector-based context representation and then adds vehicles to the map in an autoregressive manner, generating their long trajectories. The encoder captures the traffic context, including the HD road map and the states of all vehicles, which is crucial for creating a detailed and accurate representation of the traffic scenario. TrafficGen's autoregressive model iteratively encodes the current traffic context with an attention mechanism and decodes a vehicle's initial state, followed by the generation of its long trajectory. This method allows TrafficGen to produce not only static snapshots of traffic scenarios but also dynamic traffic flows with realistic vehicle behaviors over time. The generated scenarios can be used to augment existing traffic by adding new vehicles and extending fragmented trajectories, thereby enhancing the diversity and complexity of the traffic scenarios for simulation and testing. In addition, the authors in \cite{li2021scegene} propose SceGene, which is a bio-inspired GAI method that encodes traffic scenarios into genotypes and phenotypes to facilitate the generation of diverse and realistic traffic situations for robust testing. The proposed method leverages genetic algorithms for scenario generation, including crossover and mutation operations, and introduces a probabilistic-based selection algorithm and a scenario repair approach to enhance the authenticity of the generated scenarios. The authors in \cite{tan2021scenegen} propose SceneGen, a neural autoregressive model that generates traffic scenes by inserting actors with varying classes, sizes, orientations, and velocities into a scene based on the state of the ego-vehicle and a HD map, thereby improving the fidelity of simulations for self-driving vehicles.

Multi-agent trajectory generation involves simulating the future paths of multiple agents, such as vehicles, bicycles, and pedestrians, within a given scene, which is particularly relevant for developing and evaluating self-driving algorithms. The authors in \cite{guo2023scenedm} present SceneDM, a novel framework motivated by the need for realistic scene-level multi-agent motion simulations in vehicular applications, crucial for the development and evaluation of self-driving algorithms. SceneDM employs GAI through diffusion models and a Transformer-based network to generate consistent and smooth future trajectories for various agents, including vehicles, bicycles, and pedestrians, while ensuring interaction awareness and adherence to traffic rules. Furthermore, the authors in \cite{pronovost2023scenario} propose ``Scenario Diffusion," a novel diffusion-based architecture for generating synthetic traffic scenarios to validate the safety of connected and autonomous driving, motivated by the need for diverse and controllable scenario generation in vehicular applications. The proposed method combines latent diffusion, object detection, and trajectory regression to generate distributions of synthetic agent poses, orientations, and trajectories, which can be conditioned on a HD map and descriptive tokens, allowing for a high degree of control over the generated scenarios.

LLMs such as GPT-4, BERT, and PaLM have been instrumental in advancing Natural Language Processing (NLP) and are now being leveraged for traffic scenario generation in IoV. These models can understand and generate plain texts, enabling them to create complex traffic scenarios for autonomous vehicle simulations. By describing scenarios linguistically, LLMs can generate XML files for traffic simulation tools like SUMO, bypassing the need for manual or programmatic scenario creation. Therefore, the authors in \cite{bandi2023power} introduce a method that utilizes LLMs to generate these scenarios linguistically, bypassing traditional user interfaces or structured scenario files. The proposed method involves using GPT-4 to process natural language descriptions and generate XML files for the SUMO traffic simulator, with an intermediary application to ensure the correct transformation of LLM outputs into valid simulation files. In addition, the authors in \cite{tan2023language} introduce LCTGen, a GAI model designed to address the challenge of creating dynamic and realistic traffic scenarios for vehicular applications, which is crucial for the development and testing of autonomous driving systems and IoV. LCTGen generates traffic scenarios by first encoding a natural language description into a structured representation using an LLM and then using a transformer-based decoder architecture to produce the traffic scenario. The process begins with the Interpreter module, which translates the textual description into a structured representation that captures the essential elements of a scenario. This structured representation is then combined with a map retrieved from a dataset of maps to provide context for the scenario. The Generator, which is a query-based transformer model, takes this structured representation and map as input and produces the traffic scenario by modeling interactions between different agents and between agents and the map. The Generator places all agents in a single forward pass and supports end-to-end training, efficiently capturing the dynamics of the traffic scene. During inference, the Generator samples the most probable values from the predicted distributions to generate the final traffic scenario, which includes the positions, attributes, and motions of each agent over time.

\subsubsection{Scene Simulation}

Realistic video simulation within the context of IoV is crucial for developing and testing self-driving technologies. To this regard, the authors in \cite{chen2021geosim} propose GeoSim, a GAI method motivated by the need for realistic and scalable sensor simulation in vehicular applications, particularly for self-driving cars. GeoSim synthesizes novel urban driving scenarios by augmenting real images with dynamic objects and rendering them at novel poses, ensuring physical realism and high-level control in the generated images. GeoSim ensures physical realism by leveraging a geometry-aware simulation-by-composition procedure that involves realistic object placement, rendering novel views of dynamic objects from a diverse asset bank and blending the rendered image segments into existing scenes. The process begins with the construction of a 3D assets bank from sensor data, which is robust against challenges such as sparsity of observations, occlusions, and lighting changes, ensuring consistency and a strong shape prior. During simulation, GeoSim uses HD maps and LiDAR data to propose plausible locations for objects, considering the full 3D scene layout to maintain realistic behavior and interactions within the scene. Image-based rendering is then employed to handle occlusions properly, and NN-based image in-painting is used to seamlessly blend inserted objects, adjusting for color inconsistencies and removing sharp boundaries. 
Aiming at obtaining high-quality, large-scale multi-view video data for autonomous driving applications, the authors in \cite{li2023drivingdiffusion} propose DrivingDiffusion, a GAI framework that synthesizes realistic multi-view videos controlled by a 3D layout, ensuring cross-view and cross-frame consistency and enhancing instance quality through a cascaded process of multi-view image generation, temporal video generation, and post-processing. The proposed methods leverage denoising diffusion probabilistic models (DDPM) and latent diffusion models (LDM) for image synthesis, with innovations such as a consistency module and local prompt design to maintain spatial-temporal coherence and improve generation quality.  It addresses the challenges of maintaining cross-view consistency, cross-frame consistency, and high-quality instance generation by cascading multi-view single-frame image generation with single-view video generation and employing post-processing techniques. The framework ensures multi-view consistency through information exchange between adjacent cameras and employs a Key-Frame controller in the temporal model to maintain cross-frame consistency. Additionally, DrivingDiffusion introduces a local prompt to improve the quality of generated instances and uses a temporal sliding window algorithm in post-processing to enhance the cross-view consistency of subsequent frames and extend video length.

\section{Future Research Directions}

\subsection{Privacy-preserving Collaborative Inference}

Privacy-preserving collaborative inference focuses on creating methodologies for executing AI inference tasks across IoV while safeguarding individual data privacy, which can be developed based on several cryptographic techniques such as federated learning, homomorphic encryption, and secure multi-party computation. In IoV, vehicles are increasingly interconnected, generating a vast amount of data, which can be leveraged for GAI applications, such as traffic flow simulation and accident avoidance prediction, but must be handled in a way that does not compromise user privacy. Therefore, striking a balance between computational efficiency and privacy preservation is a significant challenge, especially in dynamic IoV systems with continuous data streams. However, successful implementation of privacy-preserving collaborative inference can facilitate real-time GAI applications in IoV without breaching user data privacy, hence fostering trust and wider adoption.

\subsection{Low-carbon Multimodal Perception}

Low-carbon multimodal perception is all about crafting AI models that can handle and interpret various types of data (including images, sounds, and sensor data) while keeping energy use to a minimum. This is crucial as we aim to lower the carbon footprint of our computing infrastructure. As the scale of IoV increases, the energy consumption and environmental impact of AI deployments cannot be overlooked. The goal here is to create AI models that are energy-efficient and can manage multimodal data without the need of heavy computation. This supports sustainable IoV applications like eco-friendly routing and energy-efficient vehicle operation. The trick is to design AI models that are power-saving but still highly accurate. Achieving this can lead to less environmental impact, cost savings in operations, and keeping in step with worldwide environmental goals.

\subsection{Parameter-efficient Fine-tuning of Local Experts}

Parameter-efficient fine-tuning of local experts focuses on creating effective strategies to adapt large, pre-trained AI models (our local experts) for specific tasks in IoV systems, using minimal additional training. Given the dynamic nature of IoV, it is essential that our AI models can adapt quickly to new data or scenarios, without causing a significant computational strain. The goal of parameter-efficient fine-tuning is to achieve high performance without re-training the whole model, thereby saving both computational resources and energy. The key challenge lies in reducing the parameters needed for fine-tuning, without increasing the computational load or reducing performance. In this way, it will allow for smarter, localized decision-making and the use of advanced AI capabilities on resource-limited IoV devices.

\subsection{Retrieval-augmented Generation}

Retrieval-augmented generation (RAG) can boost the capabilities of GAI models by incorporating mechanisms that retrieve and use relevant data from large datasets during the generation process. This is especially important for IoV applications like autonomous driving or traffic management, where the latest information is vital to bridge the gap between generative and retrieval-based AI to enhance the accuracy and relevance of the content generated. Although integrating retrieval mechanisms for MoE-based generative models without sacrificing speed or efficiency is a considerable challenge, successful execution could significantly improve the quality and relevance of generated outputs, which enables more accurate and context-aware applications in IoV.



\section{Conclusions}

In this survey, we introduce the integration of MoE and multimodal GAI which presents a transformative approach for advancing AGI in IoV. The potential of MoE and GAI to enhance cognitive functions and enable distributed, collaborative AI across vehicular networks has been underscored, addressing the challenges of high mobility and resource constraints. Additionally, future research directions including privacy-preserving collaborative inference and communication-efficient MoE architectural design have been identified as critical to realizing the full potential of AGI in IoV. The convergence of MoE and GAI within IoV systems is set to not only augment vehicular capabilities but also to drive innovation in intelligent transportation and autonomous vehicular technologies.

\bibliographystyle{ieeetr}
\bibliography{main}

\begin{thebibliography}{100}

\bibitem{bubeck2023sparks}
S.~Bubeck, V.~Chandrasekaran, R.~Eldan, J.~Gehrke, E.~Horvitz, E.~Kamar, P.~Lee, Y.~T. Lee, Y.~Li, S.~Lundberg, {\em et~al.}, ``Sparks of artificial general intelligence: Early experiments with gpt-4,'' {\em arXiv preprint arXiv:2303.12712}, 2023.

\bibitem{zhang2019mobile}
J.~Zhang and K.~B. Letaief, ``Mobile edge intelligence and computing for the internet of vehicles,'' {\em Proceedings of the IEEE}, vol.~108, no.~2, pp.~246--261, 2019.

\bibitem{naeem2020generative}
F.~Naeem, S.~Seifollahi, Z.~Zhou, and M.~Tariq, ``A generative adversarial network enabled deep distributional reinforcement learning for transmission scheduling in internet of vehicles,'' {\em IEEE Transactions on Intelligent Transportation Systems}, vol.~22, no.~7, pp.~4550--4559, 2020.

\bibitem{chen2023vehicle}
X.~Chen, Y.~Deng, H.~Ding, G.~Qu, H.~Zhang, P.~Li, and Y.~Fang, ``Vehicle as a service (vaas): Leverage vehicles to build service networks and capabilities for smart cities,'' {\em arXiv preprint arXiv:2304.11397}, 2023.

\bibitem{zhang2024smart}
Y.~Zhang, H.~An, Z.~Fang, G.~Xu, Y.~Zhou, X.~Chen, and Y.~Fang, ``{SmartCooper}: Vehicle collaborative perception under adaptive fusion and judger mechanism,'' in {\em 2024 IEEE International Conference on Robotics and Automation (ICRA)}, (Yokohama, Japan), IEEE, May 13-17 2024.

\bibitem{qu2024model}
K.~Qu, W.~Zhuang, Q.~Ye, W.~Wu, and X.~Shen, ``Model-assisted learning for adaptive cooperative perception of connected autonomous vehicles,'' {\em IEEE Transactions on Wireless Communications}, 2024.

\bibitem{yan2024forging}
X.~Yan, H.~Zhang, Y.~Cai, J.~Guo, W.~Qiu, B.~Gao, K.~Zhou, Y.~Zhao, H.~Jin, J.~Gao, {\em et~al.}, ``Forging vision foundation models for autonomous driving: Challenges, methodologies, and opportunities,'' {\em arXiv preprint arXiv:2401.08045}, 2024.

\bibitem{kosaraju2019social}
V.~Kosaraju, A.~Sadeghian, R.~Mart{\'\i}n-Mart{\'\i}n, I.~Reid, H.~Rezatofighi, and S.~Savarese, ``Social-bigat: Multimodal trajectory forecasting using bicycle-gan and graph attention networks,'' {\em Advances in Neural Information Processing Systems}, vol.~32, 2019.

\bibitem{khattar2019mvae}
D.~Khattar, J.~S. Goud, M.~Gupta, and V.~Varma, ``Mvae: Multimodal variational autoencoder for fake news detection,'' in {\em The world wide web conference}, pp.~2915--2921, 2019.

\bibitem{xu2023versatile}
X.~Xu, Z.~Wang, G.~Zhang, K.~Wang, and H.~Shi, ``Versatile diffusion: Text, images and variations all in one diffusion model,'' in {\em Proceedings of the IEEE/CVF International Conference on Computer Vision}, pp.~7754--7765, 2023.

\bibitem{zhang2024mm}
D.~Zhang, Y.~Yu, C.~Li, J.~Dong, D.~Su, C.~Chu, and D.~Yu, ``Mm-llms: Recent advances in multimodal large language models,'' {\em arXiv preprint arXiv:2401.13601}, 2024.

\bibitem{tang2019future}
F.~Tang, Y.~Kawamoto, N.~Kato, and J.~Liu, ``Future intelligent and secure vehicular network toward 6g: Machine-learning approaches,'' {\em Proceedings of the IEEE}, vol.~108, no.~2, pp.~292--307, 2019.

\bibitem{wen2023road}
L.~Wen, X.~Yang, D.~Fu, X.~Wang, P.~Cai, X.~Li, T.~Ma, Y.~Li, L.~Xu, D.~Shang, {\em et~al.}, ``On the road with gpt-4v (ision): Early explorations of visual-language model on autonomous driving,'' {\em arXiv preprint arXiv:2311.05332}, 2023.

\bibitem{fedus2022review}
W.~Fedus, J.~Dean, and B.~Zoph, ``A review of sparse expert models in deep learning,'' {\em arXiv preprint arXiv:2209.01667}, 2022.

\bibitem{galesic2023beyond}
M.~Galesic, D.~Barkoczi, A.~M. Berdahl, D.~Biro, G.~Carbone, I.~Giannoccaro, R.~L. Goldstone, C.~Gonzalez, A.~Kandler, A.~B. Kao, {\em et~al.}, ``Beyond collective intelligence: Collective adaptation,'' {\em Journal of the Royal Society interface}, vol.~20, no.~200, p.~20220736, 2023.

\bibitem{fedus2022switch}
W.~Fedus, B.~Zoph, and N.~Shazeer, ``Switch transformers: Scaling to trillion parameter models with simple and efficient sparsity,'' {\em The Journal of Machine Learning Research}, vol.~23, no.~1, pp.~5232--5270, 2022.

\bibitem{zhenxing2022switch}
M.~Zhenxing and D.~Xu, ``Switch-nerf: Learning scene decomposition with mixture of experts for large-scale neural radiance fields,'' in {\em The Eleventh International Conference on Learning Representations}, 2022.

\bibitem{sun2021applications}
Z.~Sun, Y.~Liu, J.~Wang, G.~Li, C.~Anil, K.~Li, X.~Guo, G.~Sun, D.~Tian, and D.~Cao, ``Applications of game theory in vehicular networks: A survey,'' {\em IEEE Communications Surveys \& Tutorials}, vol.~23, no.~4, pp.~2660--2710, 2021.

\bibitem{katare2023survey}
D.~Katare, D.~Perino, J.~Nurmi, M.~Warnier, M.~Janssen, and A.~Y. Ding, ``A survey on approximate edge ai for energy efficient autonomous driving services,'' {\em IEEE Communications Surveys \& Tutorials}, 2023.

\bibitem{mao2023green}
Y.~Mao, X.~Yu, K.~Huang, Y.-J.~A. Zhang, and J.~Zhang, ``Green edge ai: A contemporary survey,'' {\em arXiv preprint arXiv:2312.00333}, 2023.

\bibitem{wan2023efficient}
Z.~Wan, X.~Wang, C.~Liu, S.~Alam, Y.~Zheng, Z.~Qu, S.~Yan, Y.~Zhu, Q.~Zhang, M.~Chowdhury, {\em et~al.}, ``Efficient large language models: A survey,'' {\em arXiv preprint arXiv:2312.03863}, 2023.

\bibitem{mcintosh2023google}
T.~R. McIntosh, T.~Susnjak, T.~Liu, P.~Watters, and M.~N. Halgamuge, ``From google gemini to openai q*(q-star): A survey of reshaping the generative artificial intelligence (ai) research landscape,'' {\em arXiv preprint arXiv:2312.10868}, 2023.

\bibitem{yang2023llm4drive}
Z.~Yang, X.~Jia, H.~Li, and J.~Yan, ``Llm4drive: A survey of large language models for autonomous driving,'' {\em arXiv e-prints}, pp.~arXiv--2311, 2023.

\bibitem{chen2023rte}
J.~Chen, X.~Wang, and X.~S. Shen, ``Rte: Rapid and reliable trust evaluation for collaborator selection and time-sensitive task handling in internet of vehicles,'' {\em IEEE Internet of Things Journal}, 2023.

\bibitem{zhou2023vision}
X.~Zhou, M.~Liu, B.~L. Zagar, E.~Yurtsever, and A.~C. Knoll, ``Vision language models in autonomous driving and intelligent transportation systems,'' {\em arXiv preprint arXiv:2310.14414}, 2023.

\bibitem{xu2017internet}
W.~Xu, H.~Zhou, N.~Cheng, F.~Lyu, W.~Shi, J.~Chen, and X.~Shen, ``Internet of vehicles in big data era,'' {\em IEEE/CAA Journal of Automatica Sinica}, vol.~5, no.~1, pp.~19--35, 2017.

\bibitem{ye2023accuracy}
X.~Ye, K.~Qu, W.~Zhuang, and X.~Shen, ``Accuracy-aware cooperative sensing and computing for connected autonomous vehicles,'' {\em IEEE Transactions on Mobile Computing}, 2023.

\bibitem{biswas2023autonomous}
A.~Biswas and H.-C. Wang, ``Autonomous vehicles enabled by the integration of iot, edge intelligence, 5g, and blockchain,'' {\em Sensors}, vol.~23, no.~4, p.~1963, 2023.

\bibitem{zardari2022adaptive}
N.~A. Zardari, R.~Ngah, O.~Hayat, and A.~H. Sodhro, ``Adaptive mobility-aware and reliable routing protocols for healthcare vehicular network,'' {\em Mathematical Biosciences and Engineering}, vol.~19, no.~7, pp.~7156--7177, 2022.

\bibitem{jin2022mobility}
H.~Jin, P.~Zhang, H.~Dong, X.~Wei, Y.~Zhu, and T.~Gu, ``Mobility-aware and privacy-protecting qos optimization in mobile edge networks,'' {\em IEEE Transactions on Mobile Computing}, 2022.

\bibitem{ince2022real}
S.~{\.I}nce, Z.~E. Baiat, and {\c{S}}.~Baydere, ``Real-time video data traffic management for publish-subscribe based messaging system,'' in {\em 2022 International Conference on Smart Applications, Communications and Networking (SmartNets)}, pp.~1--6, IEEE, 2022.

\bibitem{liu2021blockchain}
Y.~Liu, X.~Guan, Y.~Peng, H.~Chen, T.~Ohtsuki, and Z.~Han, ``Blockchain-based task offloading for edge computing on low-quality data via distributed learning in the internet of energy,'' {\em IEEE Journal on Selected Areas in Communications}, vol.~40, no.~2, pp.~657--676, 2021.

\bibitem{khairnar2013performance}
V.~D. Khairnar and K.~Kotecha, ``Performance of vehicle-to-vehicle communication using ieee 802.11 p in vehicular ad-hoc network environment,'' {\em arXiv preprint arXiv:1304.3357}, 2013.

\bibitem{gu2020control}
P.~Gu, C.~Hua, W.~Xu, R.~Khatoun, Y.~Wu, and A.~Serhrouchni, ``Control channel anti-jamming in vehicular networks via cooperative relay beamforming,'' {\em IEEE internet of things journal}, vol.~7, no.~6, pp.~5064--5077, 2020.

\bibitem{lin2021tulvcan}
C.-H. Lin, S.-C. Lin, and E.~Blasch, ``Tulvcan: Terahertz ultra-broadband learning vehicular channel-aware networking,'' in {\em IEEE INFOCOM 2021-IEEE Conference on Computer Communications Workshops (INFOCOM WKSHPS)}, pp.~1--6, IEEE, 2021.

\bibitem{falahatraftar2021conditional}
F.~Falahatraftar, S.~Pierre, and S.~Chamberland, ``A conditional generative adversarial network based approach for network slicing in heterogeneous vehicular networks,'' in {\em Telecom}, vol.~2, pp.~141--154, MDPI, 2021.

\bibitem{seo2018gids}
E.~Seo, H.~M. Song, and H.~K. Kim, ``Gids: Gan based intrusion detection system for in-vehicle network,'' in {\em 2018 16th Annual Conference on Privacy, Security and Trust (PST)}, pp.~1--6, IEEE, 2018.

\bibitem{chen2023gan}
X.~Chen, K.~Xiao, L.~Luo, Y.~Li, and L.~Chen, ``Gan-ivds: An intrusion detection system for intelligent connected vehicles based on generative adversarial networks,'' in {\em 2023 8th International Conference on Data Science in Cyberspace (DSC)}, pp.~237--244, IEEE, 2023.

\bibitem{qiu2022unsupervised}
Y.~Qiu, T.~Misu, and C.~Busso, ``Unsupervised scalable multimodal driving anomaly detection,'' {\em IEEE Transactions on Intelligent Vehicles}, 2022.

\bibitem{monshizadeh2021improving}
M.~Monshizadeh, V.~Khatri, M.~Gamdou, R.~Kantola, and Z.~Yan, ``Improving data generalization with variational autoencoders for network traffic anomaly detection,'' {\em IEEE Access}, vol.~9, pp.~56893--56907, 2021.

\bibitem{aslam2023vae}
N.~Aslam and M.~H. Kolekar, ``A-vae: Attention based variational autoencoder for traffic video anomaly detection,'' in {\em 2023 IEEE 8th International Conference for Convergence in Technology (I2CT)}, pp.~1--7, IEEE, 2023.

\bibitem{li2022mfvt}
M.~Li, D.~Han, D.~Li, H.~Liu, and C.-C. Chang, ``Mfvt: an anomaly traffic detection method merging feature fusion network and vision transformer architecture,'' {\em EURASIP Journal on Wireless Communications and Networking}, vol.~2022, no.~1, p.~39, 2022.

\bibitem{nwafor2022canbert}
E.~Nwafor and H.~Olufowobi, ``Canbert: A language-based intrusion detection model for in-vehicle networks,'' in {\em 2022 21st IEEE International Conference on Machine Learning and Applications (ICMLA)}, pp.~294--299, IEEE, 2022.

\bibitem{jagadish2022conditional}
D.~Jagadish, A.~Chauhan, and L.~Mahto, ``Conditional variational autoencoder networks for autonomous vehicle path prediction,'' {\em Neural Processing Letters}, vol.~54, no.~5, pp.~3965--3978, 2022.

\bibitem{chen2023tasks}
J.~Chen, C.~Guo, R.~Lin, and C.~Feng, ``Tasks-oriented joint resource allocation scheme for the internet of vehicles with sensing, communication and computing integration,'' {\em China Communications}, vol.~20, no.~3, pp.~27--42, 2023.

\bibitem{raza2021task}
S.~Raza, S.~Wang, M.~Ahmed, M.~R. Anwar, M.~A. Mirza, and W.~U. Khan, ``Task offloading and resource allocation for iov using 5g nr-v2x communication,'' {\em IEEE Internet of Things Journal}, vol.~9, no.~13, pp.~10397--10410, 2021.

\bibitem{li2021adaptive}
M.~Li, J.~Gao, L.~Zhao, and X.~Shen, ``Adaptive computing scheduling for edge-assisted autonomous driving,'' {\em IEEE Transactions on Vehicular Technology}, vol.~70, no.~6, pp.~5318--5331, 2021.

\bibitem{huang2022blockchain}
C.~Huang, W.~Wang, D.~Liu, R.~Lu, and X.~Shen, ``Blockchain-assisted personalized car insurance with privacy preservation and fraud resistance,'' {\em IEEE Transactions on Vehicular Technology}, vol.~72, no.~3, pp.~3777--3792, 2022.

\bibitem{hou2023data}
J.~Hou, D.~Liu, C.~Huang, W.~Zhuang, X.~Shen, R.~Sun, and B.~Ying, ``Data protection: Privacy-preserving data collection with validation,'' {\em IEEE Transactions on Dependable and Secure Computing}, 2023.

\bibitem{huang2020dapa}
C.~Huang, R.~Lu, J.~Ni, and X.~Shen, ``Dapa: A decentralized, accountable, and privacy-preserving architecture for car sharing services,'' {\em IEEE Transactions on Vehicular Technology}, vol.~69, no.~5, pp.~4869--4882, 2020.

\bibitem{kim2023anomaly}
T.~Kim, J.~Kim, and I.~You, ``An anomaly detection method based on multiple lstm-autoencoder models for in-vehicle network,'' {\em Electronics}, vol.~12, no.~17, p.~3543, 2023.

\bibitem{nguyen2023transformer}
T.~P. Nguyen, H.~Nam, and D.~Kim, ``Transformer-based attention network for in-vehicle intrusion detection,'' {\em IEEE Access}, 2023.

\bibitem{li2020detecting}
Y.~Li, L.~Zhang, Z.~Lv, and W.~Wang, ``Detecting anomalies in intelligent vehicle charging and station power supply systems with multi-head attention models,'' {\em IEEE Transactions on Intelligent Transportation Systems}, vol.~22, no.~1, pp.~555--564, 2020.

\bibitem{shazeer2017outrageously}
N.~Shazeer, A.~Mirhoseini, K.~Maziarz, A.~Davis, Q.~Le, G.~Hinton, and J.~Dean, ``Outrageously large neural networks: The sparsely-gated mixture-of-experts layer,'' {\em arXiv preprint arXiv:1701.06538}, 2017.

\bibitem{lewis2021base}
M.~Lewis, S.~Bhosale, T.~Dettmers, N.~Goyal, and L.~Zettlemoyer, ``Base layers: Simplifying training of large, sparse models,'' in {\em International Conference on Machine Learning}, pp.~6265--6274, PMLR, 2021.

\bibitem{lepikhin2020gshard}
D.~Lepikhin, H.~Lee, Y.~Xu, D.~Chen, O.~Firat, Y.~Huang, M.~Krikun, N.~Shazeer, and Z.~Chen, ``Gshard: Scaling giant models with conditional computation and automatic sharding,'' {\em arXiv preprint arXiv:2006.16668}, 2020.

\bibitem{du2022glam}
N.~Du, Y.~Huang, A.~M. Dai, S.~Tong, D.~Lepikhin, Y.~Xu, M.~Krikun, Y.~Zhou, A.~W. Yu, O.~Firat, {\em et~al.}, ``Glam: Efficient scaling of language models with mixture-of-experts,'' in {\em International Conference on Machine Learning}, pp.~5547--5569, PMLR, 2022.

\bibitem{shen2023traffic}
Z.~E. Shen, ``Traffic volume prediction on highway network with mix-of-expert transformer,'' 2023.

\bibitem{wang2022st}
H.~Wang, J.~Chen, Z.~Fan, Z.~Zhang, Z.~Cai, and X.~Song, ``St-expertnet: A deep expert framework for traffic prediction,'' {\em IEEE Transactions on Knowledge and Data Engineering}, 2022.

\bibitem{petersen2022data}
P.~Petersen, T.~Rudolf, and E.~Sax, ``A data-driven energy estimation based on the mixture of experts method for battery electric vehicles.,'' in {\em VEHITS}, pp.~384--390, 2022.

\bibitem{yuan2023temporal}
R.~Yuan, M.~Abdel-Aty, Q.~Xiang, Z.~Wang, and X.~Gu, ``A temporal multi-gate mixture-of-experts approach for vehicle trajectory and driving intention prediction,'' {\em IEEE Transactions on Intelligent Vehicles}, 2023.

\bibitem{fraser2023deep}
B.~Fraser, A.~Perrusqu{\'\i}a, D.~Panagiotakopoulos, and W.~Guo, ``A deep mixture of experts network for drone trajectory intent classification and prediction using non-cooperative radar data,'' in {\em 2023 IEEE Symposium Series on Computational Intelligence (SSCI)}, pp.~1--6, IEEE, 2023.

\bibitem{fang2020multi}
S.~Fang and A.~Choromanska, ``Multi-modal experts network for autonomous driving,'' in {\em 2020 IEEE International Conference on Robotics and Automation (ICRA)}, pp.~6439--6445, IEEE, 2020.

\bibitem{pini2023safe}
S.~Pini, C.~S. Perone, A.~Ahuja, A.~S.~R. Ferreira, M.~Niendorf, and S.~Zagoruyko, ``Safe real-world autonomous driving by learning to predict and plan with a mixture of experts,'' in {\em 2023 IEEE International Conference on Robotics and Automation (ICRA)}, pp.~10069--10075, IEEE, 2023.

\bibitem{morra2023mixo}
L.~Morra, A.~Biondo, N.~Poerio, and F.~Lamberti, ``Mixo: Mixture of experts-based visual odometry for multicamera autonomous systems,'' {\em IEEE Transactions on Consumer Electronics}, 2023.

\bibitem{john2018estimation}
V.~John, A.~Boyali, H.~Tehrani, K.~Ishimaru, M.~Konishi, Z.~Liu, and S.~Mita, ``Estimation of steering angle and collision avoidance for automated driving using deep mixture of experts,'' {\em IEEE Transactions on Intelligent Vehicles}, vol.~3, no.~4, pp.~571--584, 2018.

\bibitem{enzweiler2011multilevel}
M.~Enzweiler and D.~M. Gavrila, ``A multilevel mixture-of-experts framework for pedestrian classification,'' {\em IEEE Transactions on Image Processing}, vol.~20, no.~10, pp.~2967--2979, 2011.

\bibitem{wang2022sfl}
Z.~Wang, P.~Sun, Y.~Hu, and A.~Boukerche, ``Sfl: A high-precision traffic flow predictor for supporting intelligent transportation systems,'' in {\em GLOBECOM 2022-2022 IEEE Global Communications Conference}, pp.~251--256, IEEE, 2022.

\bibitem{wang2022traffic}
T.~Wang, S.~Li, L.~Ji, and D.~Zhang, ``Traffic prediction based on dynamic temporal graph convolutional networks,'' in {\em Proceedings of the 2022 6th International Conference on Electronic Information Technology and Computer Engineering}, pp.~1033--1039, 2022.

\bibitem{yang2022urban}
J.~Yang, ``Urban traffic flow prediction with deep neural network,'' {\em Security and Communication Networks}, vol.~2022, 2022.

\bibitem{maity2023data}
A.~Maity and S.~Sarkar, ``Data-driven probabilistic energy consumption estimation for battery electric vehicles with model uncertainty,'' {\em International Journal of Green Energy}, pp.~1--18, 2023.

\bibitem{lu2022vehicle}
Y.~Lu, W.~Wang, X.~Hu, P.~Xu, S.~Zhou, and M.~Cai, ``Vehicle trajectory prediction in connected environments via heterogeneous context-aware graph convolutional networks,'' {\em IEEE Transactions on Intelligent Transportation Systems}, 2022.

\bibitem{yang2022study}
Z.~Yang, X.~Kang, S.~Li, C.~Zhao, J.~Zhang, and Y.~Gong, ``Study on the trajectory positioning and prediction framework of unmanned aerial vehicle based on long short-term memory neural network,'' in {\em 2022 8th International Conference on Big Data and Information Analytics (BigDIA)}, pp.~205--212, IEEE, 2022.

\bibitem{li2023towards}
X.~Li, Y.~Bai, P.~Cai, L.~Wen, D.~Fu, B.~Zhang, X.~Yang, X.~Cai, T.~Ma, J.~Guo, {\em et~al.}, ``Towards knowledge-driven autonomous driving,'' {\em arXiv preprint arXiv:2312.04316}, 2023.

\bibitem{danapal2020sensor}
G.~Danapal, G.~A. Santos, J.~P.~C. da~Costa, B.~J. Praciano, and G.~P. Pinheiro, ``Sensor fusion of camera and lidar raw data for vehicle detection,'' in {\em 2020 Workshop on Communication Networks and Power Systems (WCNPS)}, pp.~1--6, IEEE, 2020.

\bibitem{acerbo2021safe}
F.~S. Acerbo, M.~Alirczaei, H.~Van~der Auweraer, and T.~D. Son, ``Safe imitation learning on real-life highway data for human-like autonomous driving,'' in {\em 2021 IEEE International Intelligent Transportation Systems Conference (ITSC)}, pp.~3903--3908, IEEE, 2021.

\bibitem{vitelli2022safetynet}
M.~Vitelli, Y.~Chang, Y.~Ye, A.~Ferreira, M.~Wo{\l}czyk, B.~Osi{\'n}ski, M.~Niendorf, H.~Grimmett, Q.~Huang, A.~Jain, {\em et~al.}, ``Safetynet: Safe planning for real-world self-driving vehicles using machine-learned policies,'' in {\em 2022 International Conference on Robotics and Automation (ICRA)}, pp.~897--904, IEEE, 2022.

\bibitem{wylde2012safe}
M.~J. Wylde, ``Safe motion planning for autonomous driving,'' 2012.

\bibitem{cho2023dynamic}
H.~M. Cho and E.~Kim, ``Dynamic object-aware visual odometry (vo) estimation based on optical flow matching,'' {\em IEEE Access}, vol.~11, pp.~11642--11651, 2023.

\bibitem{paponpen2022implementation}
K.~Paponpen, K.~Sucharitpongpan, N.~Termsaithong, and P.~Chaipunya, ``The implementation of steering angle estimation on miniature raspberry pi-based autonomous car,'' in {\em 2022 IEEE 17th Conference on Industrial Electronics and Applications (ICIEA)}, pp.~1037--1042, IEEE, 2022.

\bibitem{gurumurthy2017deligan}
S.~Gurumurthy, R.~Kiran~Sarvadevabhatla, and R.~Venkatesh~Babu, ``Deligan: Generative adversarial networks for diverse and limited data,'' in {\em Proceedings of the IEEE conference on computer vision and pattern recognition}, pp.~166--174, 2017.

\bibitem{park2018megan}
D.~K. Park, S.~Yoo, H.~Bahng, J.~Choo, and N.~Park, ``Megan: Mixture of experts of generative adversarial networks for multimodal image generation,'' {\em arXiv preprint arXiv:1805.02481}, 2018.

\bibitem{chai2023improved}
Y.~Chai, Q.~Yin, and J.~Zhang, ``Improved training of mixture-of-experts language gans,'' in {\em ICASSP 2023-2023 IEEE International Conference on Acoustics, Speech and Signal Processing (ICASSP)}, pp.~1--5, IEEE, 2023.

\bibitem{yu2021mixture}
Q.~Yu, M.~S. Kavitha, and T.~Kurita, ``Mixture of experts with convolutional and variational autoencoders for anomaly detection,'' {\em Applied Intelligence}, vol.~51, pp.~3241--3254, 2021.

\bibitem{kopf2021mixture}
A.~Kopf, V.~Fortuin, V.~R. Somnath, and M.~Claassen, ``Mixture-of-experts variational autoencoder for clustering and generating from similarity-based representations on single cell data,'' {\em PLoS computational biology}, vol.~17, no.~6, p.~e1009086, 2021.

\bibitem{shi2019variational}
Y.~Shi, B.~Paige, P.~Torr, {\em et~al.}, ``Variational mixture-of-experts autoencoders for multi-modal deep generative models,'' {\em Advances in neural information processing systems}, vol.~32, 2019.

\bibitem{fleig2023edge}
E.~Fleig, J.~Geistert, E.~Bochinski, R.~Jongebloed, and T.~Sikora, ``Edge-aware autoencoder design for real-time mixture-of-experts image compression,'' in {\em 2023 IEEE International Symposium on Circuits and Systems (ISCAS)}, pp.~1--5, IEEE, 2023.

\bibitem{fleig2023steered}
E.~Fleig, E.~Bochinski, and T.~Sikora, ``Steered mixture-of-experts autoencoder design for real-time image modelling and denoising,'' in {\em Real-time Processing of Image, Depth and Video Information 2023}, vol.~12571, pp.~161--170, SPIE, 2023.

\bibitem{Luo2023ImageSV}
F.~Luo, J.-P. Xiang, J.~Zhang, X.~Han, and W.~Yang, ``Image super-resolution via latent diffusion: A sampling-space mixture of experts and frequency-augmented decoder approach,'' {\em ArXiv}, vol.~abs/2310.12004, 2023.

\bibitem{Mi2023SwitchNeRFLS}
Z.~Mi and D.~Xu, ``Switch-nerf: Learning scene decomposition with mixture of experts for large-scale neural radiance fields,'' in {\em International Conference on Learning Representations}, 2023.

\bibitem{xia2022pluralistic}
X.~Xia, W.~Yang, J.~Ren, Y.~Li, Y.~Zhan, B.~Han, and T.~Liu, ``Pluralistic image completion with probabilistic mixture-of-experts,'' {\em arXiv preprint arXiv:2205.09086}, 2022.

\bibitem{jiang2022gan}
S.~Jiang, Z.~Guo, S.~Zhao, H.~Wang, and W.~Jing, ``Ce-gan: A camera image enhancement generative adversarial network for autonomous driving,'' in {\em 2022 IEEE 9th International Conference on Data Science and Advanced Analytics (DSAA)}, pp.~1--6, IEEE, 2022.

\bibitem{xu2021reliability}
W.~Xu, N.~Souly, and P.~P. Brahma, ``Reliability of gan generated data to train and validate perception systems for autonomous vehicles,'' in {\em Proceedings of the ieee/cvf winter conference on applications of computer vision}, pp.~171--180, 2021.

\bibitem{zhu2017unpaired}
J.-Y. Zhu, T.~Park, P.~Isola, and A.~A. Efros, ``Unpaired image-to-image translation using cycle-consistent adversarial networks,'' in {\em Proceedings of the IEEE international conference on computer vision}, pp.~2223--2232, 2017.

\bibitem{pak2022carnet}
A.~Pak, H.~Manjunatha, D.~Filev, and P.~Tsiotras, ``Carnet: A dynamic autoencoder for learning latent dynamics in autonomous driving tasks,'' {\em arXiv preprint arXiv:2205.08712}, 2022.

\bibitem{karlsson2018data}
E.~Karlsson and N.~Mohammadiha, ``A data-driven generative model for gps sensors for autonomous driving,'' in {\em Proceedings of the 1st International Workshop on Software Engineering for AI in Autonomous Systems}, pp.~1--5, 2018.

\bibitem{arnelid2019recurrent}
H.~Arnelid, E.~L. Zec, and N.~Mohammadiha, ``Recurrent conditional generative adversarial networks for autonomous driving sensor modelling,'' in {\em 2019 IEEE Intelligent transportation systems conference (ITSC)}, pp.~1613--1618, IEEE, 2019.

\bibitem{bhattacharyya2022modeling}
R.~Bhattacharyya, B.~Wulfe, D.~J. Phillips, A.~Kuefler, J.~Morton, R.~Senanayake, and M.~J. Kochenderfer, ``Modeling human driving behavior through generative adversarial imitation learning,'' {\em IEEE Transactions on Intelligent Transportation Systems}, vol.~24, no.~3, pp.~2874--2887, 2022.

\bibitem{kuefler2017imitating}
A.~Kuefler, J.~Morton, T.~Wheeler, and M.~Kochenderfer, ``Imitating driver behavior with generative adversarial networks,'' in {\em 2017 IEEE Intelligent Vehicles Symposium (IV)}, pp.~204--211, IEEE, 2017.

\bibitem{li2023variational}
Y.~Li, S.~Mazuelas, and Y.~Shen, ``A variational learning approach for concurrent distance estimation and environmental identification,'' {\em IEEE Transactions on Wireless Communications}, 2023.

\bibitem{yi2021improving}
D.~Yi, H.~Fang, Y.~Hua, J.~Su, M.~Quddus, and J.~Han, ``Improving synthetic to realistic semantic segmentation with parallel generative ensembles for autonomous urban driving,'' {\em IEEE Transactions on Cognitive and Developmental Systems}, vol.~14, no.~4, pp.~1496--1506, 2021.

\bibitem{hua2023domain}
Y.~Hua, J.~Sui, H.~Fang, C.~Hu, and D.~Yi, ``Domain-adapted driving scene understanding with uncertainty-aware and diversified generative adversarial networks,'' {\em CAAI Transactions on Intelligence Technology}, 2023.

\bibitem{varga2023cooperative}
B.~Varga, D.~Yang, and S.~Hohmann, ``Cooperative decision-making in shared spaces: Making urban traffic safer through human-machine cooperation,'' {\em arXiv preprint arXiv:2306.14617}, 2023.

\bibitem{kloock2023coordinated}
M.~Kloock and B.~Alrifaee, ``Coordinated cooperative distributed decision-making using synchronization of local plans,'' {\em IEEE Transactions on Intelligent Vehicles}, vol.~8, no.~2, pp.~1292--1306, 2023.

\bibitem{zhang2022ai}
K.~Zhang, L.~Zhao, C.~Dong, L.~Wu, and L.~Zheng, ``Ai-tp: Attention-based interaction-aware trajectory prediction for autonomous driving,'' {\em IEEE Transactions on Intelligent Vehicles}, vol.~8, no.~1, pp.~73--83, 2022.

\bibitem{fu2021progrpgan}
T.-y. Fu and W.-C. Lee, ``Progrpgan: Progressive gan for route planning,'' in {\em Proceedings of the 27th ACM SIGKDD Conference on Knowledge Discovery \& Data Mining}, pp.~393--403, 2021.

\bibitem{ocampo2023improving}
J.~Ocampo~Jimenez and W.~Suleiman, ``Improving path planning performance through multimodal generative models with local critics,'' {\em arXiv e-prints}, pp.~arXiv--2306, 2023.

\bibitem{he2023diffusion}
H.~He, C.~Bai, K.~Xu, Z.~Yang, W.~Zhang, D.~Wang, B.~Zhao, and X.~Li, ``Diffusion model is an effective planner and data synthesizer for multi-task reinforcement learning,'' {\em arXiv preprint arXiv:2305.18459}, 2023.

\bibitem{huang2023diffusion}
S.~Huang, Z.~Wang, P.~Li, B.~Jia, T.~Liu, Y.~Zhu, W.~Liang, and S.-C. Zhu, ``Diffusion-based generation, optimization, and planning in 3d scenes,'' in {\em Proceedings of the IEEE/CVF Conference on Computer Vision and Pattern Recognition}, pp.~16750--16761, 2023.

\bibitem{ma2023physics}
L.~Ma, S.~Qu, L.~Song, Z.~Zhang, and J.~Ren, ``A physics-informed generative car-following model for connected autonomous vehicles,'' {\em Entropy}, vol.~25, no.~7, p.~1050, 2023.

\bibitem{ma2023application}
L.~Ma and S.~Qu, ``Application of conditional generative adversarial network to multi-step car-following modeling,'' {\em Frontiers in Neurorobotics}, vol.~17, p.~1148892, 2023.

\bibitem{cui2024drive}
C.~Cui, Y.~Ma, X.~Cao, W.~Ye, and Z.~Wang, ``Drive as you speak: Enabling human-like interaction with large language models in autonomous vehicles,'' in {\em Proceedings of the IEEE/CVF Winter Conference on Applications of Computer Vision}, pp.~902--909, 2024.

\bibitem{sha2023languagempc}
H.~Sha, Y.~Mu, Y.~Jiang, L.~Chen, C.~Xu, P.~Luo, S.~E. Li, M.~Tomizuka, W.~Zhan, and M.~Ding, ``Languagempc: Large language models as decision makers for autonomous driving,'' {\em arXiv preprint arXiv:2310.03026}, 2023.

\bibitem{zhou2023navgpt}
G.~Zhou, Y.~Hong, and Q.~Wu, ``Navgpt: Explicit reasoning in vision-and-language navigation with large language models,'' {\em arXiv preprint arXiv:2305.16986}, 2023.

\bibitem{wen2023dilu}
L.~Wen, D.~Fu, X.~Li, X.~Cai, T.~Ma, P.~Cai, M.~Dou, B.~Shi, L.~He, and Y.~Qiao, ``Dilu: A knowledge-driven approach to autonomous driving with large language models,'' {\em arXiv preprint arXiv:2309.16292}, 2023.

\bibitem{feng2023trafficgen}
L.~Feng, Q.~Li, Z.~Peng, S.~Tan, and B.~Zhou, ``Trafficgen: Learning to generate diverse and realistic traffic scenarios,'' in {\em 2023 IEEE International Conference on Robotics and Automation (ICRA)}, pp.~3567--3575, IEEE, 2023.

\bibitem{li2021scegene}
A.~Li, S.~Chen, L.~Sun, N.~Zheng, M.~Tomizuka, and W.~Zhan, ``Scegene: Bio-inspired traffic scenario generation for autonomous driving testing,'' {\em IEEE Transactions on Intelligent Transportation Systems}, vol.~23, no.~9, pp.~14859--14874, 2021.

\bibitem{tan2021scenegen}
S.~Tan, K.~Wong, S.~Wang, S.~Manivasagam, M.~Ren, and R.~Urtasun, ``Scenegen: Learning to generate realistic traffic scenes,'' in {\em Proceedings of the IEEE/CVF Conference on Computer Vision and Pattern Recognition}, pp.~892--901, 2021.

\bibitem{guo2023scenedm}
Z.~Guo, X.~Gao, J.~Zhou, X.~Cai, and B.~Shi, ``Scenedm: Scene-level multi-agent trajectory generation with consistent diffusion models,'' {\em arXiv preprint arXiv:2311.15736}, 2023.

\bibitem{pronovost2023scenario}
E.~Pronovost, M.~R. Ganesina, N.~Hendy, Z.~Wang, A.~Morales, K.~Wang, and N.~Roy, ``Scenario diffusion: Controllable driving scenario generation with diffusion,'' {\em arXiv preprint arXiv:2311.02738}, 2023.

\bibitem{bandi2023power}
A.~Bandi, P.~V. S.~R. Adapa, and Y.~E. V. P.~K. Kuchi, ``The power of generative ai: A review of requirements, models, input--output formats, evaluation metrics, and challenges,'' {\em Future Internet}, vol.~15, no.~8, p.~260, 2023.

\bibitem{tan2023language}
S.~Tan, B.~Ivanovic, X.~Weng, M.~Pavone, and P.~Kraehenbuehl, ``Language conditioned traffic generation,'' {\em arXiv preprint arXiv:2307.07947}, 2023.

\bibitem{chen2021geosim}
Y.~Chen, F.~Rong, S.~Duggal, S.~Wang, X.~Yan, S.~Manivasagam, S.~Xue, E.~Yumer, and R.~Urtasun, ``Geosim: Realistic video simulation via geometry-aware composition for self-driving,'' in {\em Proceedings of the IEEE/CVF conference on computer vision and pattern recognition}, pp.~7230--7240, 2021.

\bibitem{li2023drivingdiffusion}
X.~Li, Y.~Zhang, and X.~Ye, ``Drivingdiffusion: Layout-guided multi-view driving scene video generation with latent diffusion model,'' {\em arXiv preprint arXiv:2310.07771}, 2023.

\end{thebibliography}
\end{document}